\documentclass[ALICE,manyauthors]{cernphprep}

\newcommand{\upsi}{$\Upsilon$\xspace}
\newcommand{\upsone}{$\Upsilon$(1S)\xspace}
\newcommand{\upstwo}{$\Upsilon$(2S)\xspace}
\newcommand{\upsthree}{$\Upsilon$(3S)\xspace}

\newcommand{\pt}{$p_{{\mathrm T}}$\xspace}

\newcommand{\ycms}{$y_{\mathrm{cms}}$\xspace}

\newcommand{\avncoll}{$\langle N_{\mathrm{coll}} \rangle$\xspace}

\newcommand{\avTpPb}{$\langle T_{\mathrm{pPb}} \rangle$\xspace}
\newcommand{\sqrts}{$\sqrt{s_{_{\rm {NN}}}} = 8.16$~TeV\xspace}
\newcommand{\sqrtsfive}{$\sqrt{s_{_{\rm {NN}}}} = 5.02$~TeV\xspace}
\newcommand{\sqrtsPbfive}{$\sqrt{s_{_{\rm {NN}}}} = 5.02$~TeV\xspace}
\newcommand{\sqrtsPbtwo}{$\sqrt{s_{_{\rm {NN}}}} = 2.76$~TeV\xspace}
   
\newcommand{\spp}  {\ensuremath{\sqrt{s}}\xspace}

\newcommand{\RpPb}{$R_{\mathrm{pPb}}$\xspace}

\newcommand{\QpPb}{$Q_{\mathrm{pPb}}$\xspace}

\newcommand{\pPb}{$\mbox{p--Pb}$\xspace}
\newcommand{\Pbp}{$\mbox{Pb--p}$\xspace}

\newcommand{\pp}{$\mbox{pp}$\xspace}

\usepackage[comma,square,numbers,sort&compress]{natbib}
\usepackage{hyperref}
\usepackage{graphicx}  
\usepackage{dcolumn}   
\usepackage{bm}        
\usepackage{amssymb}   
\usepackage{amsfonts}
\usepackage{graphics}
\usepackage{grffile}   
\usepackage{epsfig}
\usepackage{units}
\usepackage[usenames]{color}
\usepackage[normalem]{ulem} 
\usepackage{color}
\usepackage[utf8]{inputenc}
\usepackage[T1]{fontenc}
\usepackage{subfigure}
\usepackage{lineno}
\usepackage{floatrow}
\begin{document}%

\begin{titlepage}

\PHyear{2020}
\PHnumber{243}      
\PHdate{14 April}  

\title{$\Upsilon$ production in \pPb collisions at $\sqrt{\mathbf{\textit{s}_{_{\mathrm {NN}}}}}$ = 8.16 TeV} 
\ShortTitle{$\Upsilon$ production in \pPb collisions at $\sqrt{s_{\rm NN}}$ = 8.16 TeV}  

\Collaboration{ALICE Collaboration}%
\ShortAuthor{ALICE Collaboration}       
\begin{abstract}
\upsi production in \pPb interactions is studied at the centre-of-mass energy per nucleon--nucleon collision \sqrts with the ALICE detector at the CERN LHC. 
The measurement is performed reconstructing bottomonium resonances via their dimuon decay channel, in the centre-of-mass rapidity intervals $2.03<y_{\rm{cms}}<3.53$ and $-4.46<y_{\rm{cms}}<-2.96$, down to zero transverse momentum. 
In this work, results on the \upsone production cross section as a function of rapidity and transverse momentum are presented. The corresponding nuclear modification factor shows a suppression of the \upsone yields with respect to \mbox{pp} collisions, both at forward and backward rapidity. This suppression is stronger in the low transverse momentum region and shows no significant dependence on the centrality of the interactions. Furthermore, the \upstwo nuclear modification factor is evaluated, suggesting a suppression similar to that of the \upsone. A first measurement of the \upsthree has also been performed. Finally, results are compared with previous ALICE measurements in \pPb collisions at \sqrtsfive and with theoretical calculations.

\end{abstract}
\end{titlepage}
\setcounter{page}{2}
%
\section{Introduction}

Quarkonium resonances, i.e. bound states of a heavy quark ($\rm{Q}$) and anti-quark ($\overline{\rm{Q}}$), are well-known probes of the formation of a quark--gluon plasma (QGP) which can occur in heavy-ions collisions. The high colour-charge density reached in such a medium can, in fact, screen the binding force between the $\rm{Q}$ and $\overline{\rm{Q}}$, leading to a temperature-dependent melting of the quarkonium states according to their binding energies~\cite{Matsui:1986dk}.

A suppression of bottomonium resonances, the bound states formed by  $\rm{b}$ and $\overline{\rm{b}}$ quarks, was observed in \mbox{Pb--Pb} collisions, at the LHC energies of \sqrtsPbtwo and \sqrtsPbfive by the ALICE~\cite{Abelev:2014nua,Acharya:2018mni} and CMS~\cite{Chatrchyan:2011pe,Chatrchyan:2012lxa,Sirunyan:2018nsz} experiments.  All the \upsi resonances show a reduction in their production yields compared to \mbox{pp} interactions at the same centre-of-mass energy, scaled by the number of nucleon--nucleon collisions. Furthermore, the magnitude of the suppression is significantly different for the three resonances and it increases from the tightly bound \upsone to the loosely bound \upsthree~\cite{Chatrchyan:2011pe,Chatrchyan:2012lxa,Sirunyan:2018nsz}, as expected in a sequential suppression scenario, with the binding energies of the \upsi states ranging between $\sim$1 GeV for the \upsone to $\sim$0.2~GeV for the \upsthree~\cite{Satz:2006kba}.
Modifications to the bottomonium production might also be induced by cold nuclear matter (CNM) mechanisms not related to the formation of the QGP.
The modification of the quark and gluon structure functions for nucleons inside nuclei, modeled either via nuclear parton distribution functions (nPDFs)~\cite{Eskola:2009uj,Eskola:2016oht,Kovarik:2015cma,Kusina:2017gkz} or through a Color Glass Condensate effective theory~\cite{Fujii:2013gxa}, or the coherent energy loss of the $\rm{Q\overline{Q}}$ pair during its path through the cold nuclear medium~\cite{Arleo:2012rs} are examples of CNM effects which can influence quarkonium production~\cite{Albacete:2017qng}. 
The size of these effects is usually assessed in proton--nucleus collisions. These interactions also allow for the investigation of additional final state mechanisms, which can modify the production in particular of the more loosely bound resonances~\cite{Ferreiro:2018wbd,Ferreiro:2012rq,Du:2018wsj}. 

ALICE has published results on the modification of the \upsone production yields as a function of the centre-of-mass rapidity (\ycms) using the 2013 \pPb collisions data sample  at \sqrtsfive \cite{Abelev:2014oea}. The size of the observed suppression was found to be similar in the forward and backward  rapidity regions. Theoretical calculations based on the aforementioned CNM mechanisms fairly describe the forward-\ycms measurements, while they slightly overestimate the results obtained at backward rapidity. Furthermore, the measurement of the \upstwo to  \upsone ratio~\cite{Abelev:2014oea}, \upstwo/\upsone, was consistent, albeit within large uncertainties, with the one obtained in \pp collisions~\cite{Abelev:2014qha}, suggesting CNM effects of the same size 
on the two resonances both at forward and backward rapidity.
Consistent results were also obtained by the LHCb experiment~\cite{Aaij:2014mza} in a similar kinematic region. However, it should be noted that ATLAS~\cite{Aaboud:2017cif} and CMS~\cite{Chatrchyan:2013nza} measurements of \upstwo/\upsone  at midrapidity suggest a stronger suppression of the \upstwo with respect to the \upsone state, as expected if final state effects are at play~\cite{Ferreiro:2018wbd}.

In 2016, the LHC delivered \pPb collisions at \sqrts. The increase both in integrated luminosity, about a factor of 2 larger than the one collected in 2013, and in the bottomonium production cross section, due to the higher centre-of-mass energy, allows a more detailed study of the production of the \upsi states. In this paper, results on the \upsone production as a function of \ycms, transverse momentum (\pt)  and centrality of the collisions will be discussed and compared with the measurements performed in \pPb collisions at \sqrtsfive and with theoretical calculations. A comparison of the \upstwo and \upsthree to \upsone production yields and nuclear modification factors, integrated over \ycms, \pt and centrality, will also be presented. Finally, the results will be compared with the corresponding measurements obtained by LHCb at the same energy~\cite{Aaij:2018scz}.
It should be noted that all the presented results refer to the \upsi inclusive production, i.e. to \upsi either produced directly or coming from the feed-down of higher-mass excited states. 

\section{Experimental apparatus and data sample}

A detailed description of the ALICE apparatus and performance can be found in ~\cite{Aamodt:2008zz,Abelev:2014ffa}. The forward muon spectrometer~\cite{MS_TDR} is the main detector used in this analysis. It consists of five tracking stations made of two planes of Cathode Pad Chambers each, followed by two trigger stations each one composed by two planes of Resistive Plate Chambers. A 10 interaction-length ($\lambda_{\rm I}$) absorber, placed in front of the tracking system, filters out most of the hadrons produced in the collisions. Low-momentum muons and hadrons escaping the first absorber are stopped by a second 7.2 $\lambda_{\rm I}$-thick iron wall, placed in front of the trigger stations.  The momentum of the particles is evaluated by measuring their curvature in a dipole magnet with a 3 T$\times$m integrated field.
The muon spectrometer measures muons in the pseudorapidity interval $-4 < \eta < -2.5$ in the laboratory reference frame. It also provides single and unlike- or like-sign dimuon triggers based on the detection in the trigger system of one or two muons, respectively, having a transverse momentum higher than a programmable threshold set to $p_{{\mathrm T},\mu} = 0.5$ GeV/$c$. This threshold is not sharp and the single muon trigger efficiency reaches a plateau value of $\sim$98\% at about  $p_{{\mathrm T},\mu}\sim$1.5~GeV/$c$.

The primary interaction vertex of the collision is reconstructed using the two innermost layers of the Inner Tracking System (Silicon Pixel Detector, SPD)~\cite{Aamodt:2010aa}, extending over the pseudorapidity intervals $|\eta| < 2$ and $|\eta| < 1.4$, respectively. The V0 detector~\cite{Abbas:2013taa}, composed of two sets of scintillators covering the pseudorapidity intervals $2.8 < \eta < 5.1$ and $-3.7 < \eta < -1.7$, provides the luminosity measurement, which can also be obtained independently from the information of the T0 Cherenkov detectors~\cite{Bondila:2005xy}, covering the regions  $4.6 < \eta < 4.9$ and $-3.3 < \eta < -3$. The V0 detector is also used to provide the minimum bias (MB) trigger, defined by the coincidence of signals in the two sets of scintillators. The trigger condition used in this analysis is based on the coincidence of the MB trigger with the unlike-sign dimuon one ($\mu\mu$-$\rm{MB}$). The removal of beam-induced background is based on the timing information provided by the V0 and by two sets of Zero Degree Calorimeters (ZDC)~\cite{ALICE:2012aa} placed at $\pm$112.5~m from the interaction point, along the beamline. The ZDCs are also used for the centrality estimation as it will be discussed in Sec.~\ref{sec:analysis}. 
Finally, for the study of the \upsi production as a function of the centrality of the collisions, pile-up events in which two or more interactions occur in the same colliding bunch are removed using the information from SPD and V0. 

Further selection criteria, commonly adopted in the ALICE quarkonium analyses (see e.g.~\cite{Acharya:2018kxc,Abelev:2014oea}), are applied to the muon tracks forming the dimuon pair. Muon tracks must have a pseudorapidity value in the range $-4 < \eta_{\rm{\mu}} < -2.5$, corresponding to the muon spectrometer acceptance, and they should point to the interaction vertex to remove fake tracks and particles not directly produced in beam--beam interactions. Their transverse coordinate at the end of the front absorber ($R_{\rm{abs}}$) must be within $17.6$~cm  $< R_{\rm{abs}} < 89.5$~cm, to remove muons not passing the homogeneous region of the absorber. Finally, tracks reconstructed in the tracking chambers of the muon spectrometer should match the track segments reconstructed in the trigger system. This matching request helps to further reject hadron contamination and ensures that the reconstructed muons fulfill the trigger condition.

The data were collected with two beam configurations obtained by inverting the directions of the proton and Pb beams circulating inside the LHC. In this way it was possible to cover both a forward ($2.03 < y_{\mathrm{cms}} < 3.53$) and a backward ($-4.46 < y_{\mathrm{cms}} < -2.96$) dimuon rapidity interval, where the positive (negative) \ycms refers to the proton (Pb) beam going
towards the muon spectrometer. 
The collected integrated luminosities for the corresponding data samples, referred to as \pPb (forward rapidity) and \Pbp (backward rapidity) in the following, are $\mathcal{L}_{\rm {int}}^{\rm{pPb}} = 8.4 \pm 0.2$~{\rm nb}$^{-1}$ and $\mathcal{L}_{\rm {int}}^{\rm{Pbp}} = 12.8 \pm 0.3$~{\rm nb}$^{-1}$~\cite{ALICE-PUBLIC-2018-002}. 

\section{Data analysis}
\label{sec:analysis}

The results presented in this paper are based on an analysis procedure similar to the one described in ~\cite{Abelev:2014oea} for the study  of the \upsi production in \pPb collisions at \sqrtsfive. 

The \upsone, \upstwo and \upsthree production cross sections, corrected by the branching ratio for the decay in a muon pair ($\rm{B.R.}_{\rm{\Upsilon}\rightarrow\rm{\mu^{+}\mu^{-}}}$), are obtained, for a given ($\Delta y_{\rm cms}, \Delta p_{\rm T}$) interval, as
\begin{equation}
\frac{\rm{d}^2\sigma^\Upsilon_{\rm pPb}}{\rm{d}\it{y}_{\rm cms}\rm{d}\it{p}_{\rm T}} =  \frac{N_{\Upsilon}}
{\mathcal{L}_{\rm int}^{\rm pPb}\times(A\times\epsilon) \times\Delta y_{\rm cms}\times\Delta p_{\rm T}\times{\rm {B.R.}}_{\Upsilon\rightarrow\mu^{+}\mu^{-}}},
\label{eq:cs}
\end{equation}
where $N_{\rm{\Upsilon}}$ is the number of signal counts and ($A\times\epsilon$) is the corresponding acceptance and efficiency correction in the kinematic bin under study, while the branching ratios are ($2.48 \pm 0.05$)\% for \upsone, ($1.93 \pm 0.17$)\% for \upstwo and ($2.18 \pm 0.21$)\% for \upsthree ~\cite{Tanabashi:2018oca}. 

\begin{figure}[h!]
\begin{center}
\includegraphics[width=0.49\linewidth]{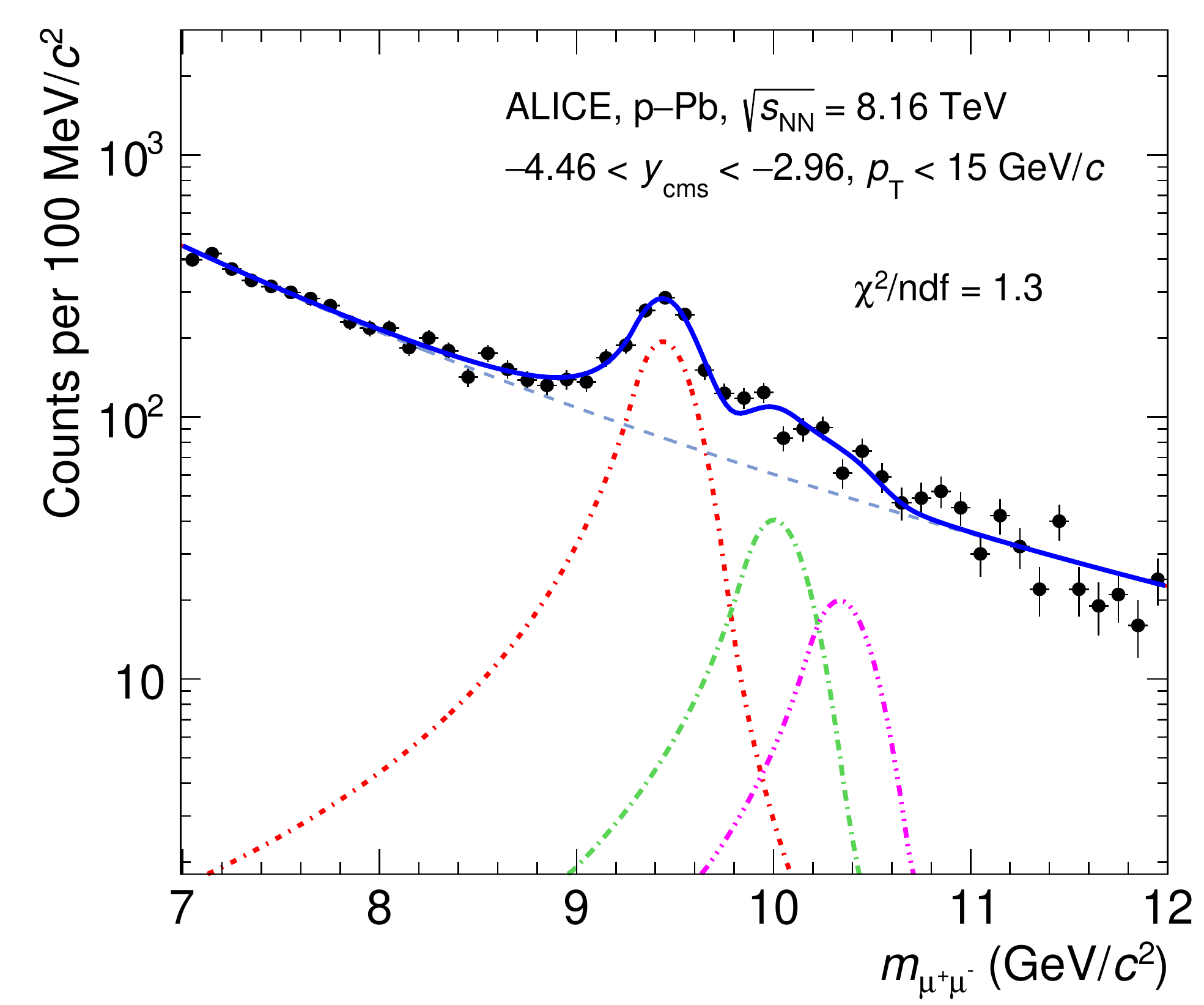}
\includegraphics[width=0.49\linewidth]{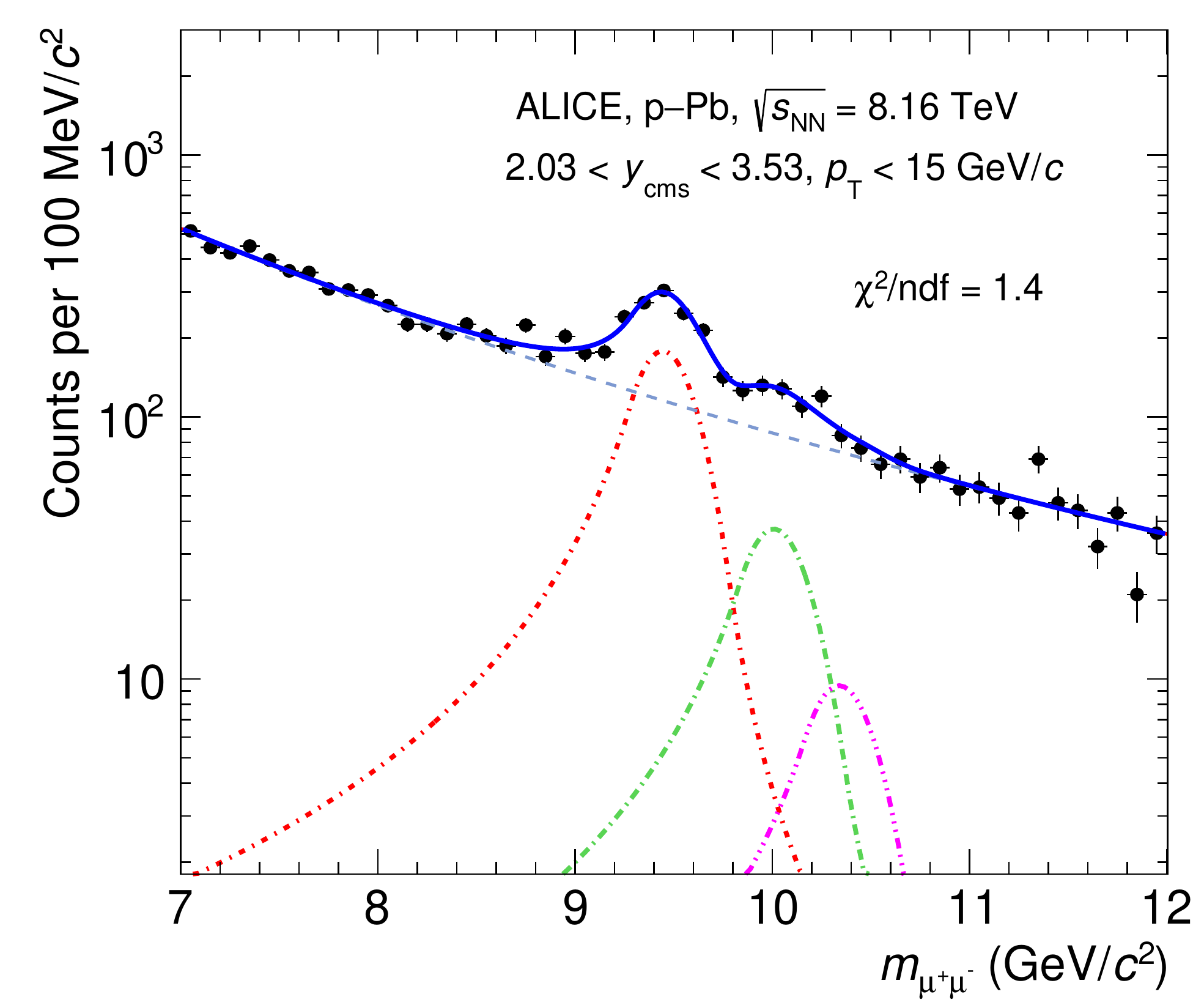}  
\caption{Invariant mass spectra of unlike-sign dimuons, integrated over \pt, for \Pbp (left panel) and \pPb (right panel) collisions. The shapes of the \upsone, \upstwo and \upsthree resonances are shown (dash-dotted lines), together with the background function (dashed line) and the total fit (solid line).}
\label{fig:invmass} 
\end{center}
\end{figure}

The number of $\Upsilon\rm{(nS)}$ is obtained by fitting the unlike-sign dimuon invariant mass spectrum with a combination of signal shapes to describe the \upsi resonances and an empirical function to model the background.
More in detail, the background is described by several combinations of exponential and polynomial functions or by a Gaussian function with a mass-dependent width. For the resonance shapes, extended Crystal Ball functions~\cite{ALICE:quarkonium}, with power-law tails on the right and left sides of the mass peak are used. Alternatively, pseudo-Gaussian functions with a mass-dependent width are also adopted~\cite{ALICE:quarkonium}. The same signal shape is chosen for all the \upsi states. 
The mass of the \upsone and its width  $\sigma_{\rm{\Upsilon\rm{(1S)}}}$ are free parameters of the fit, while 
the mass and the width of the \upstwo and \upsthree states are bound to those of the \upsone in the following way: $m_{\rm{\Upsilon\rm{(nS)}}} = m_{\rm{\Upsilon\rm{(1S)}}}+ (m_{\rm{\Upsilon\rm{(nS)}}}^{\rm{PDG}} - m_{\rm{\Upsilon\rm{(1S)}}}^{\rm{PDG}})$ and 
$\sigma_{\rm{\Upsilon\rm{(nS)}}} = \sigma_{\rm{\Upsilon\rm{(1S)}}}\times \sigma_{\rm{\Upsilon\rm{(nS)}}}^{\rm{MC}}/\sigma_{\rm{\Upsilon\rm{(1S)}}}^{\rm{MC}}$.  The mass value $m_{\rm{\Upsilon\rm{(nS)}}}^{\rm{PDG}}$ is taken from \cite{Tanabashi:2018oca} and $\sigma_{\rm{\Upsilon\rm{(nS)}}}^{\rm{MC}}$ is the width of the resonance as evaluated from a fit, with the aforementioned signal functions, to the spectrum obtained from the Monte Carlo (MC) simulation also used for the ($A\times\epsilon$) correction.
Due to the signal-over-background ratio of the order of $\sim$0.7 ($\sim$1) in \pPb (\Pbp), measured in a 3$\sigma$ region around the \upsone mass, the non-Gaussian tails of the extended Crystal Ball function can not be  kept as free parameters of the fits. Hence, they are tuned on \mbox{pp} data at \spp = 13 TeV, the largest data sample collected by ALICE so far, or, alternatively, on \pPb or \pp MC simulations at \sqrts and \spp = 8 TeV, respectively. The same tails are adopted for the \upstwo and \upsthree mass shapes.
Examples of the fit to the invariant mass spectrum, for both the \pPb and \Pbp samples, are shown in Fig.~\ref{fig:invmass}.

The number of \upsi candidates, $N_{\Upsilon}$, is evaluated as the average of the values obtained by varying the signal and background functions as well as the fitting intervals (6~ GeV/${c^2}$ $< m_{\rm{\mu\mu}}<$ 13~GeV/${c^2}$ or 7~GeV/${c^2}$ $< m_{\rm{\mu\mu}}<$ 12~GeV/${c^2}$). The statistical uncertainties are calculated as the average of the statistical uncertainties over the various fits and the standard deviation of the  distribution of the $N_{\Upsilon}$ values provides the systematic uncertainties on the signal extraction. For the \upstwo and \upsthree cases, an additional contribution to the systematic uncertainty is included, to account for possible variations of their width with respect to that of the \upsone.
In particular, their widths are allowed to vary between a minimum value $\sigma_{\rm{\Upsilon\rm{(1S)}}}$ and 
a maximum value $\sigma_{\rm{\Upsilon\rm{(1S)}}} \times \sigma_{\rm{\Upsilon\rm{(nS)}}}^{\rm{MC}}/\sigma_{\rm{\Upsilon\rm{(1S)}}}^{\rm{MC}}$, where the ratio $\sigma_{\rm{\Upsilon\rm{(nS)}}}^{\rm{MC}}/\sigma_{\rm{\Upsilon\rm{(1S)}}}^{\rm{MC}}$ is obtained from MC simulations alternative to the ones used for the  ($A\times\epsilon$) correction, i.e. based on different \upsi kinematic input shapes, as it will be discussed later on. A further 5\% systematic uncertainty is also included to account for possible residual discrepancies between the detector resolution in MC and in the data.

The total number of \upsone, integrated over the full kinematic range, amounts to $N_{\rm{\Upsilon\rm{(1S)}}} = 909 \pm 62\, \rm{(stat.)} \pm 58\, \rm{(syst.)}$ and $N_{\rm{\Upsilon\rm{(1S)}}} = 918 \pm 55\, \rm{(stat.)} \pm 51\, \rm{(syst.)}$ for the forward and backward-rapidity regions, respectively. 
Corresponding values for \upstwo are $N_{\rm{\Upsilon\rm{(2S)}}} = 192 \pm 39\,\rm{(stat.)} \pm 17\,  \rm{(syst.)}$ and $N_{\rm{\Upsilon\rm{(2S)}}} = 194 \pm 34\, \rm{(stat.)} \pm 16\,  \rm{(syst.)}$, while for the \upsthree the values are $N_{\rm{\Upsilon\rm{(3S)}}} = 48 \pm 36\,\rm{(stat.)} \pm 8\,  \rm{(syst.)}$ and $N_{\rm{\Upsilon\rm{(3S)}}} = 95 \pm 30\, \rm{(stat.)} \pm 12\,  \rm{(syst.)}$. The systematic uncertainty, amounting to $\sim$6\% for the \upsone and $\sim$8\% for the \upstwo, is dominated by the choice of the tail parameters  in the fit functions and, in the \upstwo case, also by the allowed range of variation for the $\sigma_{\rm{\Upsilon\rm{(2S)}}}$. In the \upsthree case, the systematic uncertainties are slightly larger, amounting to $\sim$17\% at forward rapidity and $\sim$12\% at backward rapidity.
For \pt- or \ycms-differential \upsone  studies, the systematic uncertainties have a similar size, reaching  $\sim$15\% only in the highest \pt bin (8~GeV/$c$ $< p_{\mathrm{T}} <$ 15~GeV/$c$).

The acceptance and efficiency correction is calculated in a  MC simulation, based on the GEANT3 transport code~\cite{GEANT3}. The MC simulation is performed on a run-by-run basis to closely follow the evolution of the performance of the detectors during the data taking. The \upsone are generated using rapidity and transverse momentum distributions tuned on \pPb or \Pbp data at \sqrts, through an iterative procedure~\cite{Acharya:2018kxc}.  
The \pt and \ycms integrated ($A\times\epsilon$) amounts to $0.300 \pm 0.006$ for the \upsone at forward rapidity and $0.273 \pm 0.007$ at backward rapidity, where the quoted uncertainties are systematic, the statistical uncertainties being negligible. The lower ($A\times\epsilon$) values measured in the \Pbp period, with respect to the \pPb one, are due to detector instabilities which affected temporarily the behaviour of two tracking chambers. The limited size of the data sample do not allow for a similar tuning of the \pt and \ycms distributions on data for the \upstwo and \upsthree resonances, hence the same shapes as for the \upsone are used. The resulting ($A\times\epsilon$) values show a negligible difference with respect to the \upsone ones.
The systematic uncertainties on ($A\times\epsilon$) include contributions related to the choice of the MC \pt and \ycms input distributions for the  \upsi states and to the evaluation of the tracking and trigger efficiencies. The systematic uncertainties associated to the MC \upsi input shapes are evaluated as the maximum difference between the ($A\times\epsilon$) evaluated with the aforementioned MC tuned on data and the values extracted from alternative MC samples based on \pt and \ycms \upsi distributions either measured by the LHCb experiment in \pp collisions at $\sqrt{s}$ = 8 TeV~\cite{Aaij:2015awa} or obtained from existing CDF and LHC \pp measurements~\cite{Acosta:2001gv,LHCb:2012aa,Khachatryan:2010zg} via a procedure similar to the one described in~\cite{Bossu:2011qe}. Nuclear shadowing is also included to account for its  influence on the bottomonium kinematic distributions. These systematic uncertainties 
amount to 1\% (1.3\%) for \pPb and 1\% (1.6\%) for \Pbp for the three resonances. They have a negligible \pt-dependence, while they reach up to 4\% at the edges of the rapidity intervals. 
The systematic uncertainty on the trigger efficiency consists of two contributions, one related to the evaluation of the intrinsic efficiency of each muon-trigger chamber (1\%) and one to small differences between the trigger response function estimated via data and MC (0.6\% in  \pPb and 0.2\% in \Pbp, when integrating over \ycms and \pt). This last source of uncertainty is below 1\% also for the \pt or \ycms-differential studies. 
The systematic uncertainty associated to the tracking efficiency is evaluated  comparing the dimuon tracking efficiencies computed both in data and MC. These efficiencies are computed combining the efficiency of each single muon-tracking chamber, obtained relying on the redundancy of the tracking system. The resulting systematic uncertainties amount to 1\% for  \pPb and 2\% for \Pbp, for both the \ycms and \pt differential studies and for results integrated  over the kinematic domain. Finally, an additional 1\% systematic uncertainty on the choice of the $\chi^2$ cut on the matching between the tracks reconstructed in the tracking and in the trigger systems is included. The systematic uncertainties associated to the trigger, tracking and matching efficiencies are considered to be identical for both the \upsone and \upstwo resonances. 

The integrated luminosities are obtained as $\mathcal{L}_{\rm {int}} = N_{\rm{MB}}/\sigma_{\rm MB}$. The number of equivalent minimum bias events, $N_{\rm MB}$, is evaluated by multiplying the number of events collected with the $\mu\mu$-$\rm{MB}$ trigger by a factor $F_{\rm{norm}}$, corresponding to the inverse of the probability of having a triggered dimuon in a MB event~\cite{Acharya:2018kxc}. 
This quantity is computed, run by run, as the ratio between the number of collected MB triggers and the number of times the dimuon trigger condition is verified in the MB trigger sample.
Once averaged over all the runs, considering as weight the number of $\mu\mu$-$\rm{MB}$ triggers in each run, $F_{\rm{norm}}$ amounts to $679 \pm 7$ at forward rapidity and $372 \pm 4$ at backward rapidity. The quoted uncertainty (1\%) is systematic and accounts for differences coming from an alternative evaluation method, based on the information provided by the level-0 trigger scalers, as detailed in~\cite{Abelev:2013yxa}. The V0-based MB cross section ($\sigma_{\rm MB}$) is measured from a van der Meer scan, and it amounts to $2.09 \pm 0.04$ b for the \pPb configuration and $2.10 \pm 0.04$ b for the \Pbp one~\cite{ALICE-PUBLIC-2018-002}. In the luminosity systematic uncertainty quoted in Table~\ref{tab:syst}, the uncertainties on  $F_{\rm{norm}}$ and $\sigma_{\rm MB}$ are combined, together with a 1.1\% (0.6\%) contribution due to the difference between the luminosities obtained with the V0 and T0 detectors in the \pPb (\Pbp) configurations~\cite{ALICE-PUBLIC-2018-002}.

The nuclear effects on the \upsi production are studied comparing the corresponding \pPb production cross section to the one measured in \pp collisions,  $\rm{d}^2\sigma^\Upsilon_{\rm pp}/\rm{d}\it{y}_{\rm cms}\rm{d}\it{p}_{\rm T}$, obtained at the same centre-of-mass energy and scaled by the atomic mass number of the Pb nucleus ($A_{\rm Pb} = 208$), through the so-called nuclear modification factor \RpPb, defined as
\begin{equation}
R_{\rm{pPb}} = \frac{\rm{d}^2\sigma^\Upsilon_{\rm pPb}/\rm{d}\it{y}_{\rm cms}\rm{d}\it{p}_{\rm T}}{A_{\rm Pb} \times \rm{d}^2\sigma^\Upsilon_{\rm pp}/\rm{d}\it{y}_{\rm cms}\rm{d}\it{p}_{\rm T}}.
\label{eq:RpPb}
\end{equation}
The proton--proton reference is based on the LHCb measurements of the bottomonium production cross section in \pp collisions at $\sqrt{s}$ = 8 TeV~\cite{Aaij:2015awa}, in $-4.5 < y_{\mathrm{cms}} < -2.5$ and $2 < y_{\mathrm{cms}} < 4$, corrected by a factor to account for the slightly different centre-of-mass energies of the interactions. This correction factor is evaluated interpolating the LHCb measurements at $\sqrt{s} =$ 7, 8 and 13 TeV~\cite{Aaij:2015awa,Aaij:2018pfp}, as detailed in ~\cite{ALICE:upsilon}. It amounts to $1.02$ for both the \upsone and \upstwo, showing a negligible \ycms dependence and varying by 1\% from low to high \pt. A systematic uncertainty on the determination of this factor (1\%) is assigned, based on the choice of the  different functions used for the energy-interpolation. 
The \upsi production cross sections in \pp collisions at  $\sqrt{s}$ = 8 TeV are also measured by ALICE~\cite{Adam:2015rta}. The results show good agreement with the corresponding LHCb values, but unlike the LHCb measurements, they cover a slightly narrower rapidity region, $2.5 < y_{\mathrm{cms}} < 4$, which does not match the rapidity coverage of the \pPb measurements.
The $\sigma^{\rm{\Upsilon(1S)}}_{\rm{pp}}$ cross sections, integrated over \pt and \ycms, are $98.5 \pm 0.1 \; \rm{(stat.)} \pm 3.4 \; \rm{(syst.)} \; {\rm nb}$ in the range $2.03 < y_{\mathrm{cms}} < 3.53$ and $62.0 \pm 0.1 \; \rm{(stat.)} \pm 2.1  \;\rm{(syst.)} \; {\rm nb}$ in the range  $-4.46 < y_{\mathrm{cms}} < -2.96$. The corresponding cross sections for the \upstwo are about a factor 3 smaller, being  $\sigma^{\rm{\Upsilon(2S)}}_{\rm{pp}} = 31.9 \pm 0.1\; \rm{(stat.)} \pm 2.9 \;\rm{(syst.)}\; {\rm  nb}$ at forward rapidity and  $19.7 \pm 0.05\; \rm{(stat.)} \pm 1.8 \;\rm{(syst.)}\;  {\rm nb}$ at backward rapidity. The \upsthree production cross sections are $\sigma^{\rm{\Upsilon(3S)}}_{\rm{pp}} = 12.9 \pm 0.1\; \rm{(stat.)} \pm 1.3 \;\rm{(syst.)}\; {\rm  nb}$ at forward rapidity and  $8.3 \pm 0.1\; \rm{(stat.)} \pm 0.8 \;\rm{(syst.)}\;  {\rm nb}$ at backward rapidity.

The large data sample collected in \pPb collisions in 2016 allows the \upsone production also to be studied as a function of the collision centrality. The centrality determination is based on a hybrid model, as discussed in detail in~\cite{Adam:2014qja}. In this approach, the centrality is determined by measuring the energy released in the ZDC positioned in the Pb-going direction.  For each ZDC-selected centrality class, the average number of collisions \avncoll is obtained as $\langle N_{\mathrm{coll}} \rangle = \langle N_{\rm{part}} \rangle$-1, assuming  the charged particle multiplicity measured at midrapidity is proportional to the number of participant nucleons, $N_{\rm{part}}$.  
The centrality classes used in this analysis correspond to 2--20\%, 20--40\%, 40--60\% and 60--90\% of the MB cross section. The 0--2\% most central collisions are excluded from this analysis because the fraction of events coming from pile-up in the ZDC is large in this centrality interval and a residual contamination might still be present in spite of the applied pile-up rejection cuts~\cite{Adam:2015jsa}.

For centrality studies, the modification induced by the nuclear matter on the \upsone production is quantified through the nuclear modification factor denoted by \QpPb, to be distinguished from \RpPb since potential biases from the centrality estimation, unrelated to nuclear effects, might be present~\cite{Adam:2014qja}. The \QpPb is defined as

\begin{equation}
Q_{\rm{pPb}} = \frac{N_{\rm{\Upsilon}}}{\rm{B.R.}_{\rm{\Upsilon}\rightarrow\rm{\mu^{+}\mu^{-}}} \times \mathit{N}_{\mathrm{MB}} \times (\mathit{A}\times\epsilon) \times \langle \mathit{T}_{\mathrm{pPb}} \rangle \times \sigma^{\rm{\Upsilon}}_{\mathrm{pp}}}.
\label{eq:QpPb}
\end{equation}

The quantities entering Eq.~\ref{eq:QpPb} are evaluated according to the previously discussed procedure, with few minor differences. When extracting the \upsone signal, for example, 
no significant variation of the \upsone width as a function of the collision centrality is foreseen. Hence for centrality studies, the \upsone width is fixed to the value obtained in the fit to the centrality-integrated invariant mass spectrum. The uncertainty associated to the choice of the width is accounted for in the evaluation of the systematic uncertainty on the signal extraction. 
No significant centrality dependence is expected for the ($A\times\epsilon$) either, so the centrality-integrated values are also used for all the centrality classes.
To evaluate the number of MB events in each centrality class $i$, $F_{\rm{norm}}^i$ is obtained from the centrality-integrated quantity scaled by the ratio of the number of minimum bias and dimuon-triggered events in each centrality interval with respect to the corresponding centrality integrated quantities, $(N_{\rm {MB}}^{i}/N_{\rm {MB}})/(N_{\mu\mu-\rm{MB}}^{i}/N_{\mu\mu-\rm{MB}})$. 
Alternatively, $F_{\rm{norm}}^i$ is computed directly for each centrality class and a further 1\% difference between the two approaches is included in the systematic uncertainty. The statistical uncertainty on $F_{\rm{norm}}^{i}$ is negligible.
Finally, $\langle T_{\rm{pPb}} \rangle$ is the centrality-dependent average nuclear thickness function, computed with the Glauber framework~\cite{Miller:2007ri,Adam:2014qja}.

The systematic uncertainties entering the cross section and nuclear modification factor evaluation are summarised in Table ~\ref{tab:syst}. 

\begin{table}[h]
\centering
\begin{tabular}{c|c|c|c|c|c|c}
\hline
Sources  & \multicolumn{2}{c|}{\upsone}  &  \multicolumn{2}{c}{\upstwo}  &  \multicolumn{2}{c}{\upsthree}\\
\hline
  & \mbox{p--Pb}  &  \mbox{Pb--p} & \mbox{p--Pb}  &  \mbox{Pb--p} & \mbox{p--Pb}  &  \mbox{Pb--p}\\
\hline
\hline
Signal extraction & 6.4 (5.1--15.9) & 5.7 (5.5--8.5)  & 8.8 & 8.4 & 17.4 & 12.6\\
\hline
Trigger efficiency (II) & 1.2 (1.1--1.3)  &  1.0 (1.0--1.1)  & 1.2 & 1.0 & 1.2 & 1.0\\
\hline
Tracking efficiency (II) & 1.0  & 2.0  & 1.0  & 2.0 & 1.0 & 2.0\\
\hline
Matching efficiency (II) & 1.0  & 1.0  & 1.0  & 1.0 & 1.0 & 1.0\\
\hline
MC inputs & 1.0 (0.5--4.0)  & 1.0 (0.4--4.0) & 1.3 & 1.6 & 1.4 & 1.8\\
\hline
pp reference (II) &  0.2 (0.1--0.4) & 0.2 (0.1--0.4) & 0.2 & 0.3 & 0.2 & 0.2\\
\hline
pp reference (I,II) &  \multicolumn{2}{c|}{2.8} & \multicolumn{2}{c|}{2.8}& \multicolumn{2}{c}{2.8}\\
\hline
$\mathcal{L}_{\rm {int}}^{\rm pPb}$ (II) & 2.1  & 2.2 & 2.1  & 2.2  & 2.1  & 2.2\\
\hline
$\mathcal{L}_{\rm {int}}^{\rm pPb}$ (I,II) & 0.5 & 0.7& 0.5 & 0.7 & 0.5 & 0.7\\
\hline
Pile-up & 2.0  & 2.0 & \multicolumn{2}{c|}{--}  & \multicolumn{2}{c}{--}\\
\hline 
\avTpPb & \multicolumn{2}{c|}{2.1--5.8} &  \multicolumn{2}{c|}{--}&  \multicolumn{2}{c}{--}\\ 
\hline
B.R. (I) &  \multicolumn{2}{c|}{2.0} & \multicolumn{2}{c|}{8.8}& \multicolumn{2}{c}{9.6}\\
\hline
\end{tabular}
\caption{Systematic uncertainties, in percentage, on the three \upsi cross sections and nuclear modification factors for both p--Pb and Pb--p collisions. Ranges in parentheses refer to the maximum variation as a function of centrality, \ycms or \pt. When no ranges are specified, the quoted values are valid for both the integrated and the differential measurements. Error type I means that the uncertainties are correlated over \pt or \ycms, while error type II refers to uncertainties correlated versus centrality. If no error type is specified, the uncertainties are considered as uncorrelated. The uncertainties on the pp reference and luminosity result from the combination of  \ycms-uncorrelated and  correlated contributions. For the systematic uncertainty on the luminosity determination, the two terms, defined according to~\cite{ALICE-PUBLIC-2018-002}, are separately quoted in the table, but combined when results are shown in the figures. Uncertainties on the B.R. are taken from~\cite{Tanabashi:2018oca}.
}
\label{tab:syst}
\end{table}

When \RpPb is computed as a function of \pt or \ycms, the systematic uncertainties on the signal extraction, tracking, trigger and matching efficiencies, MC input shapes and a fraction of the uncertainty on the \pp reference are considered as bin-by-bin uncorrelated. On the contrary, the correlated contributions to the \mbox{pp} reference and the luminosity uncertainties, which are common to the \pPb or \Pbp systems, are considered as correlated over \pt or \ycms. 
In the \QpPb evaluation, the uncertainties on signal extraction, on the MC input shapes and on \avTpPb  depend on the centrality of the collision, while the other uncertainties are common to all classes and, therefore, considered as correlated over centrality. 
Even if most central events are not included in this analysis, a further 2\% centrality-uncorrelated systematic uncertainty is assigned to the \QpPb values, to account for residual pile-up which might still introduce a bias in the measurement. This systematic uncertainty is evaluated by comparing the expected pile-up fraction, computed from the pile-up probability associated to the observed interaction rate,  and the amount of pile-up events removed by the event selection procedure.
For the \upstwo and \upsthree studies, similar values of the systematic uncertainties are  obtained, the main difference being the larger signal extraction uncertainties.

\section{Results}

The inclusive \upsone production cross sections are evaluated in the rapidity regions $2.03 < y_{\rm{cms}} < 3.53$ and $-4.46 < y_{\rm{cms}} < -2.96$ and their values, computed according to Eq.~\ref{eq:cs}, are:

\begin{align*}
\sigma_{\rm{pPb}}^{\rm{\Upsilon{(1S)}}} (2.03 < y_{\rm{cms}} < 3.53) = 14.5 \pm 1.0 \; \rm{(stat.)} \pm 1.0 \;\rm{(uncor.\,syst.)} \pm 0.3 \;\rm{(cor.\,syst.)} \;\mu{\rm b}, \\
\sigma_{\rm{pPb}}^{\rm{\Upsilon{(1S)}}} (-4.46 < y_{\rm{cms}} < -2.96) = 10.5 \pm 0.6   \;\rm{(stat.)} \pm 0.7 \;\rm{(uncor.\,syst.)} \pm 0.2 \;\rm{(cor.\,syst.)}  \;\mu{\rm b}. 
\end{align*}

The corresponding values for the \upstwo production cross sections are: 
\begin{align*}
\sigma_{\rm{pPb}}^{\rm{\Upsilon{(2S)}}} (2.03 < y_{\rm{cms}} < 3.53) = 3.9 \pm 0.8 \;\rm{(stat.)} \pm 0.4 \;\rm{(uncor.\,syst.)} \pm 0.3 \;\rm{(cor.\,syst.)} \;\mu{\rm b}, \\
\sigma_{\rm{pPb}}^{\rm{\Upsilon{(2S)}}} (-4.46 < y_{\rm{cms}} < -2.96) = 2.8 \pm 0.5 \;\rm{(stat.)} \pm 0.3 \;\rm{(uncor.\,syst.)} \pm 0.3 \;\rm{(cor.\,syst.)} \; \mu{\rm b}, 
\end{align*}
and for the \upsthree are:
\begin{align*}
\sigma_{\rm{pPb}}^{\rm{\Upsilon{(3S)}}} (2.03 < y_{\rm{cms}} < 3.53) = 0.87 \pm 0.66 \;\rm{(stat.)} \pm 0.15 \;\rm{(uncor.\,syst.)} \pm 0.08 \;\rm{(cor.\,syst.)}  \;\mu{\rm b}, \\
\sigma_{\rm{pPb}}^{\rm{\Upsilon{(3S)}}} (-4.46 < y_{\rm{cms}} < -2.96) = 1.24 \pm 0.39 \;\rm{(stat.)} \pm 0.15 \;\rm{(uncor.\,syst.)} \pm 0.12 \;\rm{(cor.\,syst.)} \; \mu{\rm b}. 
\end{align*}

The systematic uncertainties have two terms, one correlated and one uncorrelated as a function of rapidity.

\begin{figure}[hpb]
\begin{center}
\includegraphics[width=0.7\linewidth]{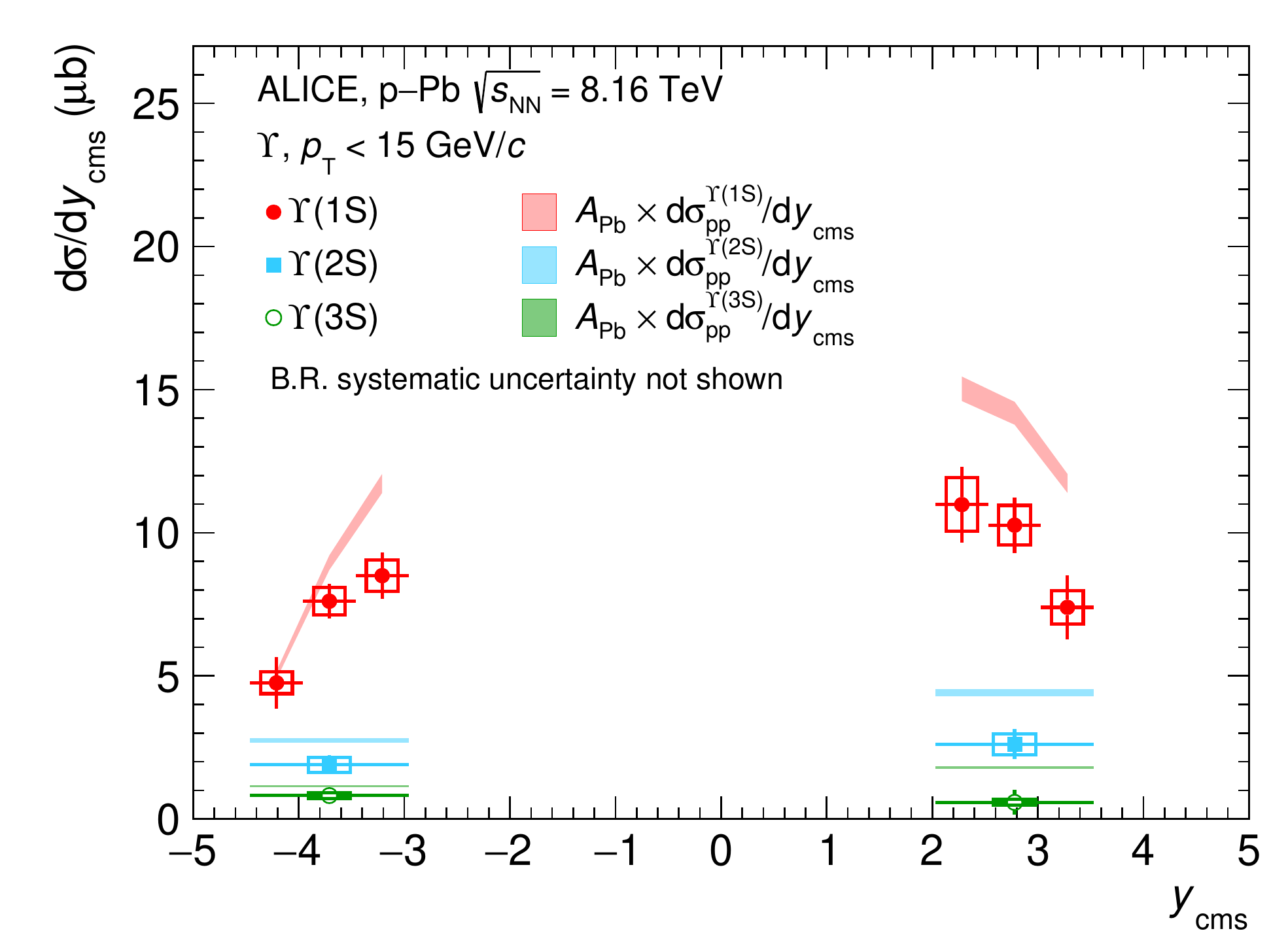} 
\caption{\upsone, \upstwo and \upsthree differential cross sections as a function of \ycms in \pPb collisions at \sqrts. 
The corresponding \pp reference cross sections, obtained through the procedure described in Sec.~\ref{sec:analysis} and scaled by $A_{\rm Pb}$, are shown as bands.}
\label{fig:cs_y} 
\end{center}
\end{figure}

\begin{figure}[htpb]
\begin{center}
\includegraphics[width=0.7\linewidth]{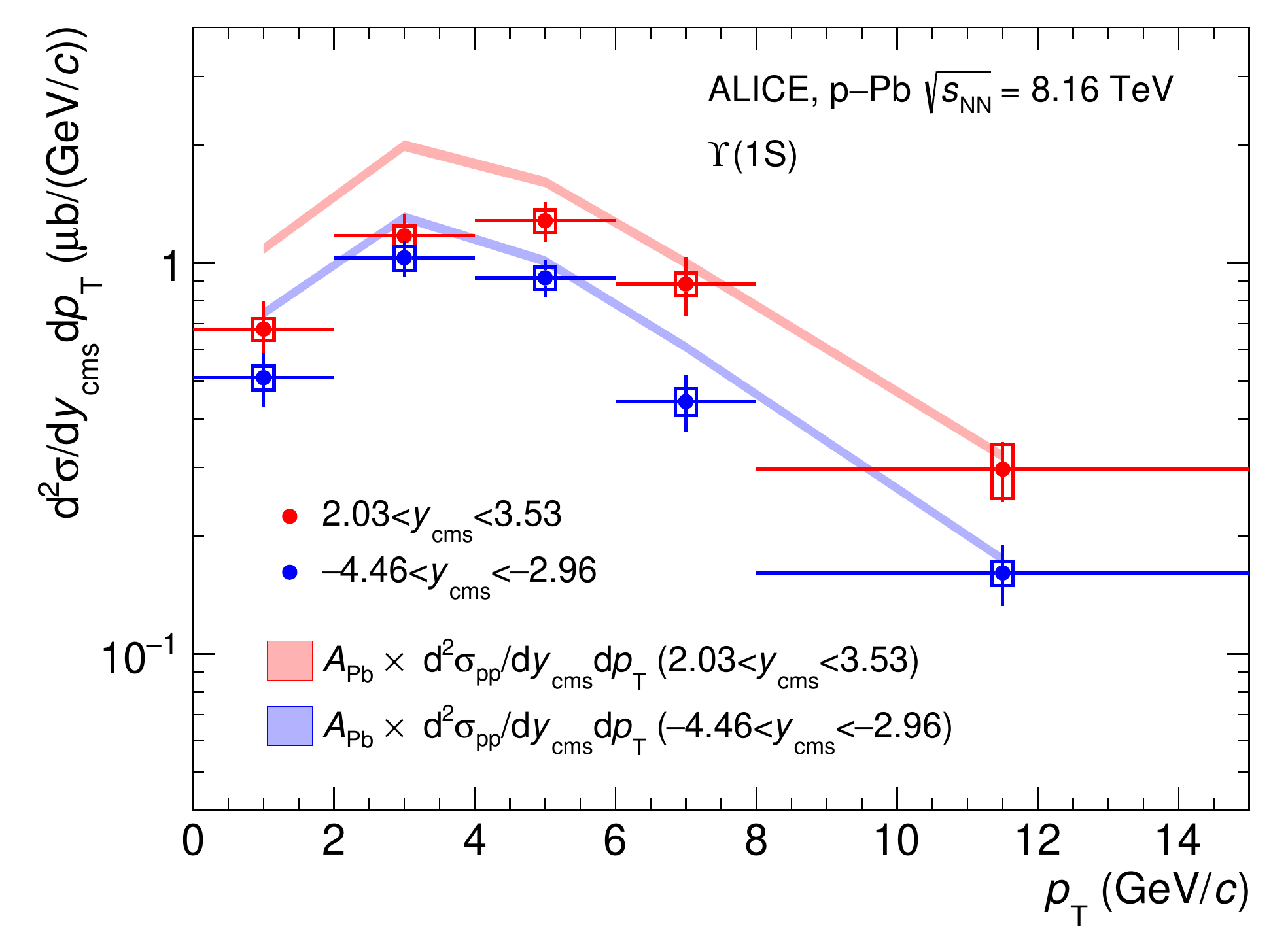} 
\caption{\upsone differential cross section as a function of \pt, at forward (closed symbols) and backward (open symbols) rapidity, at \sqrts. 
The \pp reference cross section, obtained through the procedure described in Sec.~\ref{sec:analysis} and scaled by $A_{\rm Pb}$, is shown as a band.}
\label{fig:cs_pt} 
\end{center}
\end{figure}

The data collected in \pPb collisions at \sqrts allow for the measurement of the \upsone production cross sections differentially in \ycms bins or in \pt intervals, up to \pt$<$ 15~GeV/$\it{c}$. The resulting cross sections are shown in Fig.~\ref{fig:cs_y} as a function of rapidity, integrated over  transverse momentum, and in Fig.~\ref{fig:cs_pt}, as a function of \pt, in the forward- and backward-rapidity regions. In these figures, as in all the following ones, the statistical uncertainties are shown as vertical error bars, while the systematic uncertainties are represented as boxes around the points. The horizontal error bars correspond to the \ycms or \pt bin widths.
The cross sections evaluated at forward and backward rapidities are compared with the \pp ones, obtained through the aforementioned interpolation procedure, scaled by the Pb atomic mass number.
The comparison shows that in the forward-rapidity region the \upsone cross sections are smaller than the \pp ones, in particular at low \pt, suggesting the presence of CNM effects at play in \pPb collisions. On the contrary, in the backward-rapidity range the \pp and the \pPb cross sections are closer and nuclear effects seem to have a less prominent role.

The limited available data sample allows for the evaluation of the \upstwo and \upsthree cross sections in the forward and backward-rapidity regions only integrating over the corresponding \ycms and \pt ranges, as shown in Fig.~\ref{fig:cs_y}. A suppression with respect to the corresponding  \pp reference cross sections, scaled by  $A_{\rm Pb}$, is observed.


Given the relatively small mass difference between the \upsone and \upstwo (or \upsthree) resonances, most of the systematic uncertainties, except those on the signal extraction and on the choice of the \pt- and \ycms-input shapes used in the MC, cancel in the ratio of the resonance yields, multiplied by their branching ratios, defined as
\begin{center}
${[\Upsilon\mathrm{(nS)}/\Upsilon\mathrm{(1S)}]}_{\mathrm{pPb}} = 
\frac{N_{\Upsilon\mathrm{(nS)}}/(\mathit{A}\times\epsilon)_{\Upsilon\mathrm{(nS)}}}
{N_{\Upsilon\mathrm{(1S)}}/(\mathit{A}\times\epsilon)_{\Upsilon\mathrm{(1S)}}}.$
\end{center}

\begin{figure}[ht!]
\begin{center}
\includegraphics[width=0.7\linewidth]{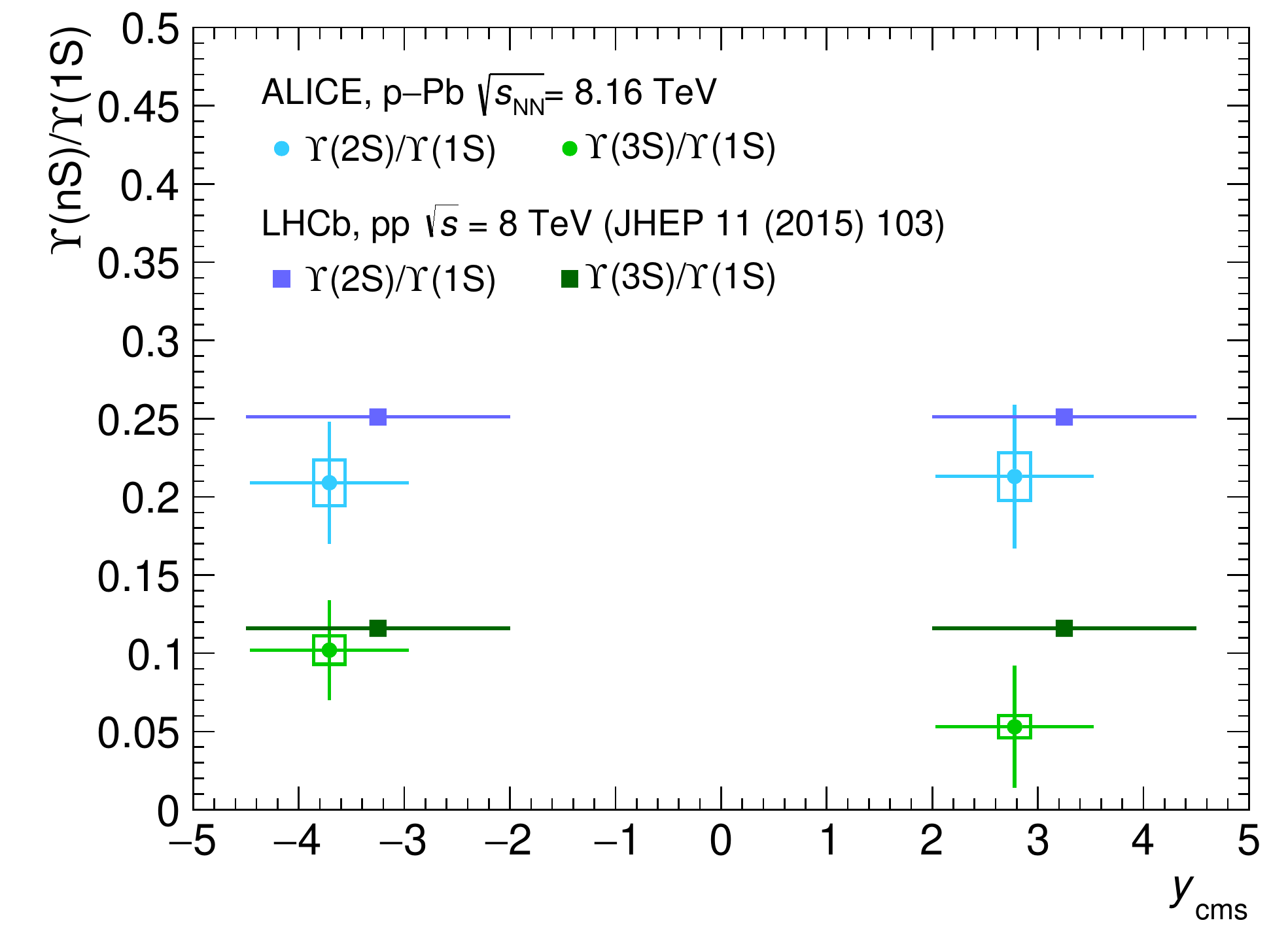}
\caption{Ratio of $\Upsilon\rm{(nS)}$ over \upsone yields in \pPb collisions at \sqrts and in \pp collisions at $\sqrt{s}$ = 8 TeV~\cite{Aaij:2015awa}. }
\label{fig:Upsilon2S1S} 
\end{center}
\end{figure}

The values of the \upstwo over \upsone ratio, obtained at forward and backward rapidity, are similar:
\begin{align*}
[\Upsilon\rm{(2S)}/\Upsilon\rm{(1S)}]_{\rm{pPb}} (2.03 < {\it y}_{\rm{cms}} < 3.53) = 0.21 \pm 0.05 \; \rm{(stat.)} \pm 0.02 \;\rm{(syst.)}, \\
[\Upsilon\rm{(2S)}/\Upsilon\rm{(1S)}]_{\rm{pPb}} (-4.46 < {\it y}_{\rm{cms}} < -2.96) = 0.21 \pm 0.04 \;\rm{(stat.)} \pm 0.01 \;\rm{(syst.)}. 
\end{align*}
As shown in Fig.~\ref{fig:Upsilon2S1S}, the ratio $[\Upsilon\rm{(2S)}/\Upsilon\rm{(1S)}]_{\rm{pPb}}$  at \sqrts is compatible, within uncertainties, with the results obtained by the LHCb Collaboration in \pp collisions at $\sqrt{s}$ = 8 TeV~\cite{Aaij:2015awa}, in a slightly wider kinematic range ($2 < y_{\rm{cms}} < 4.5$, \pt$ < 15 $~GeV/$c$). 

Similar conclusions can be obtained from the comparison of the \upsthree over \upsone ratio, also shown in Fig.~\ref{fig:Upsilon2S1S}. The corresponding values at forward and backward rapidity are:
\begin{align*}
[\Upsilon\rm{(3S)}/\Upsilon\rm{(1S)}]_{\rm{pPb}} (2.03 < {\it y}_{\rm{cms}} < 3.53) =  0.053 \pm 0.039\; \rm{(stat.)} \pm  0.007\;\rm{(syst.)}, \\
[\Upsilon\rm{(3S)}/\Upsilon\rm{(1S)}]_{\rm{pPb}} (-4.46 < {\it y}_{\rm{cms}} < -2.96) =  0.102 \pm  0.032 \;\rm{(stat.)} \pm 0.009 \;\rm{(syst.)}. 
\end{align*}

The size of nuclear effects in \pPb collisions can be better quantified through the nuclear modification factor defined in Eq.~\ref{eq:RpPb}. The numerical values for the \upsone \RpPb in the forward- and in the backward-rapidity regions, integrating over \pt, are:

\begin{align*}
R_{\rm{pPb}}^{\rm{\Upsilon(1S)}} (2.03 < y_{\rm{cms}} < 3.53) = 0.71 \pm 0.05 \;\rm{(stat.)} \pm 0.05 \;\rm{(uncor.\,syst.)}  \pm 0.02 \;\rm{(cor.\,syst.)},\\
R_{\rm{pPb}}^{\rm{\Upsilon(1S)}} (-4.46 < y_{\rm{cms}} < -2.96) = 0.81 \pm 0.05 \;\rm{(stat.)} \pm 0.05 \;\rm{(uncor.\,syst.) \pm 0.02 \;\rm{(cor.\,syst.)}}, 
\end{align*}

where (uncor. syst.) and (cor. syst.) refer to uncorrelated and correlated systematic uncertainties as a function of rapidity.

\begin{figure}[htpb]
\begin{center}
\includegraphics[width=0.7\linewidth]{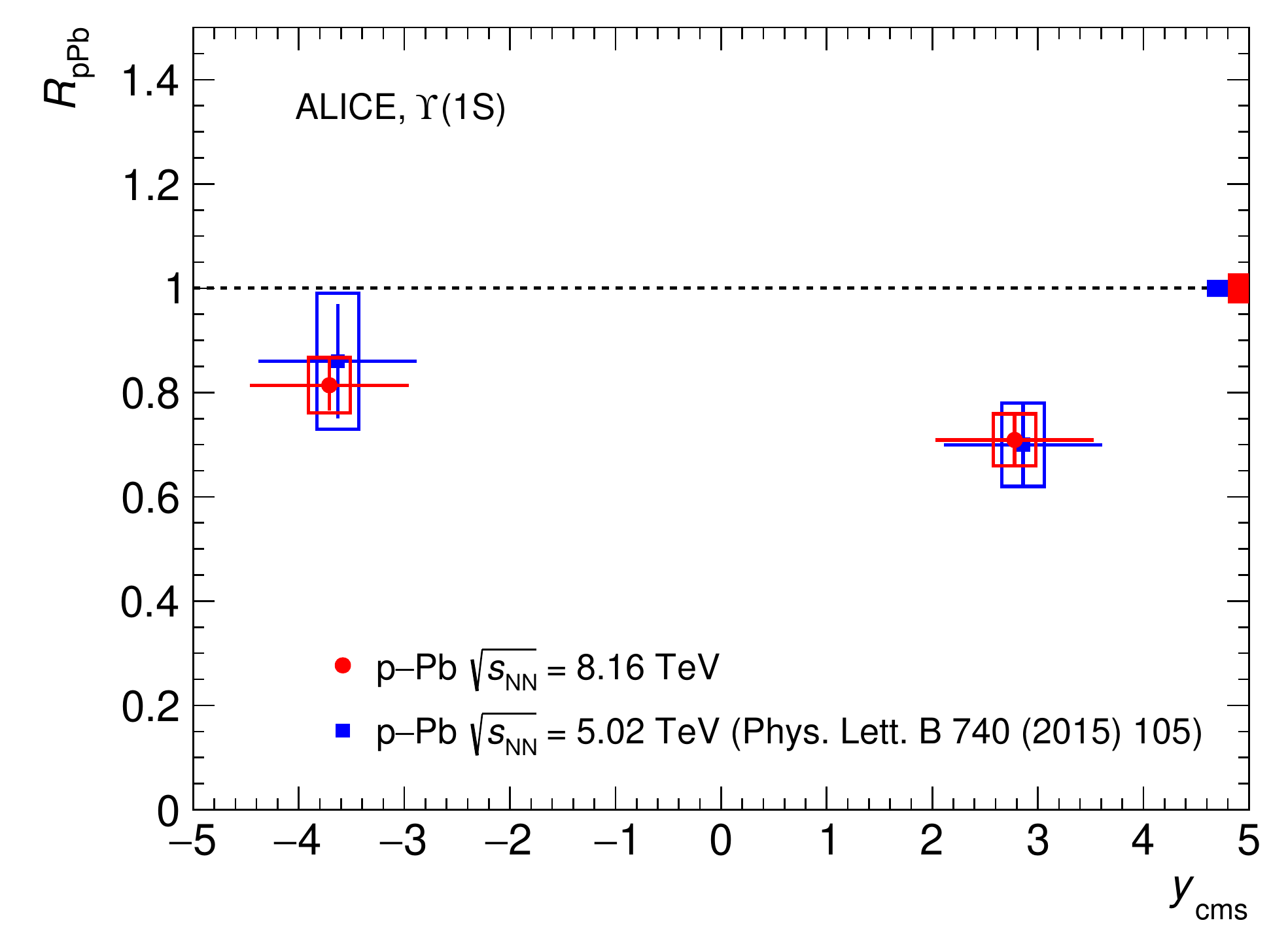} 
\caption{\upsone \RpPb values at \sqrts  compared to those obtained at \sqrtsfive in the same \ycms interval~\cite{Abelev:2014oea}. 
All systematic uncertainties are considered as uncorrelated between the results at \sqrts and \sqrtsfive. 
The \RpPb values at the two energies are slightly displaced horizontally to improve visibility.}
\label{fig:RpA} 
\end{center}
\end{figure}

The measured \RpPb values, shown in Fig.~\ref{fig:RpA},  indicate a suppression of the \upsone production in \mbox{p--Pb} collisions, with respect to the one in \pp collisions, both at forward and backward rapidity, with a slightly stronger suppression at forward \ycms. The \RpPb is found to be 4.0$\sigma$ and 2.4$\sigma$ below unity in \pPb and \Pbp collisions, respectively.  
The results are compatible with the corresponding \RpPb values measured in \mbox{p--Pb} collisions at \sqrtsfive ~\cite{Abelev:2014oea}, also shown in Fig.~\ref{fig:RpA}. From the comparison between the results obtained at the two energies, an improvement in the precision of the \upsone \RpPb measurements at \sqrts can be noticed, given the reduced size of the statistical and systematic uncertainties. The improvement of the latter contribution is mainly related to the reduction in the uncertainties associated to the tracking efficiencies and to refinements in the determination of the \pp reference~\cite{Abelev:2014oea}. 

\begin{figure}[ht!]
\begin{center}
\includegraphics[width=0.7\linewidth]{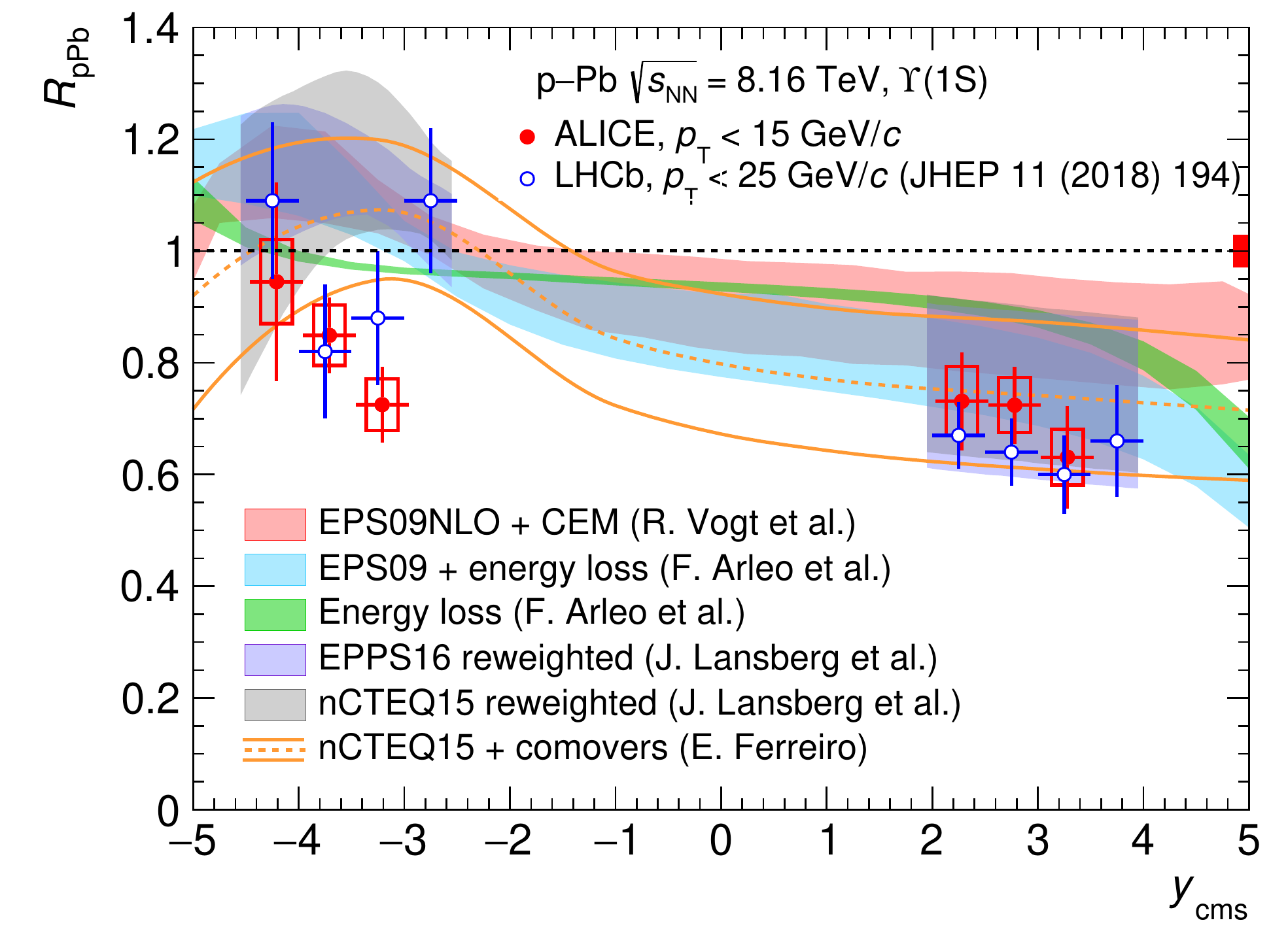} 
\caption{\upsone \RpPb values at \sqrts compared with the corresponding LHCb results~\cite{Aaij:2018scz}, as a function of \ycms.  The \RpPb values are also compared to model calculations based on several implementations of nuclear shadowing (EPS09 NLO~\cite{Vogt:2015uba,Albacete:2017qng, Eskola:2009uj}, EPPS16 and nCTEQ15~\cite{Lansberg:2016deg,Kusina:2017gkz,Shao:2015vga,Shao:2012iz,Kovarik:2015cma,Eskola:2016oht}) and on parton coherent energy loss predictions, with or without the inclusion of the EPS09 shadowing contribution ~\cite{Arleo:2012rs,Albacete:2017qng}. A theoretical model including a shadowing contribution based on nCTEQ15 nPDFs on top of a suppression induced by comover interactions~\cite{Ferreiro:2018vmr,Ferreiro:2018wbd} is also shown. 
For the LHCb results, the vertical error bars represent the quadratic sum of the statistical and systematic uncertainties.}
\label{fig:RpA_y} 
\end{center}
\end{figure}

The rapidity dependence of the \upsone \RpPb, explored in narrower \ycms intervals, is shown in Fig.~\ref{fig:RpA_y}, confirming the suppression already observed in the \ycms-integrated case. 
The results are also compared with the \upsone LHCb measurements~\cite{Aaij:2018scz} at the same centre-of-mass energy and in slightly wider kinematic ranges ($-4.5 < y_{\rm{cms}} < -2.5$ and $2 < y_{\rm{cms}} < 4$, \pt$ < 25 $~GeV/$c$). Fair agreement between the two sets of results can be seen.

\begin{figure}[ht!]
\begin{center}
\includegraphics[width=0.49\linewidth]{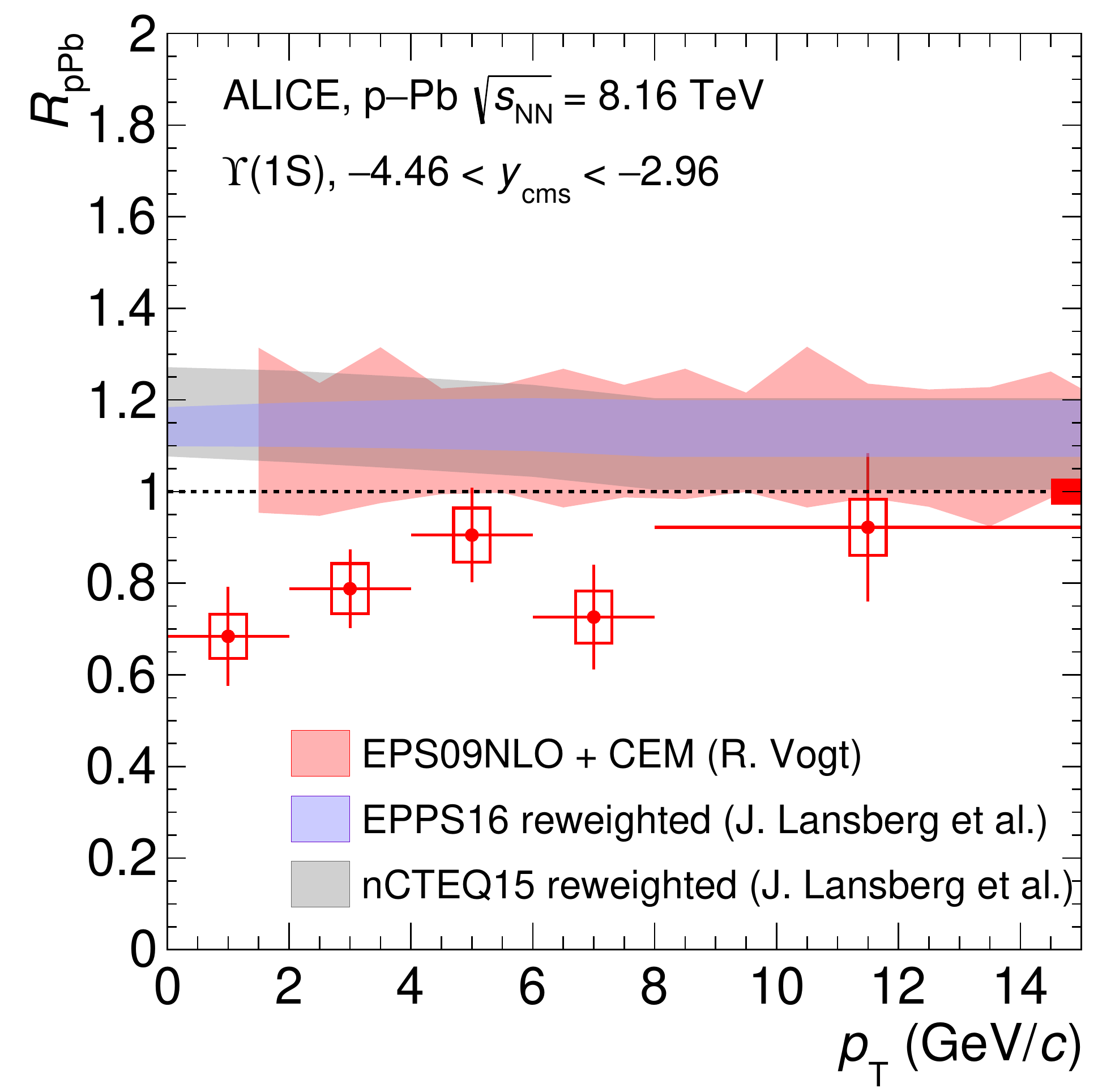}  
\includegraphics[width=0.49\linewidth]{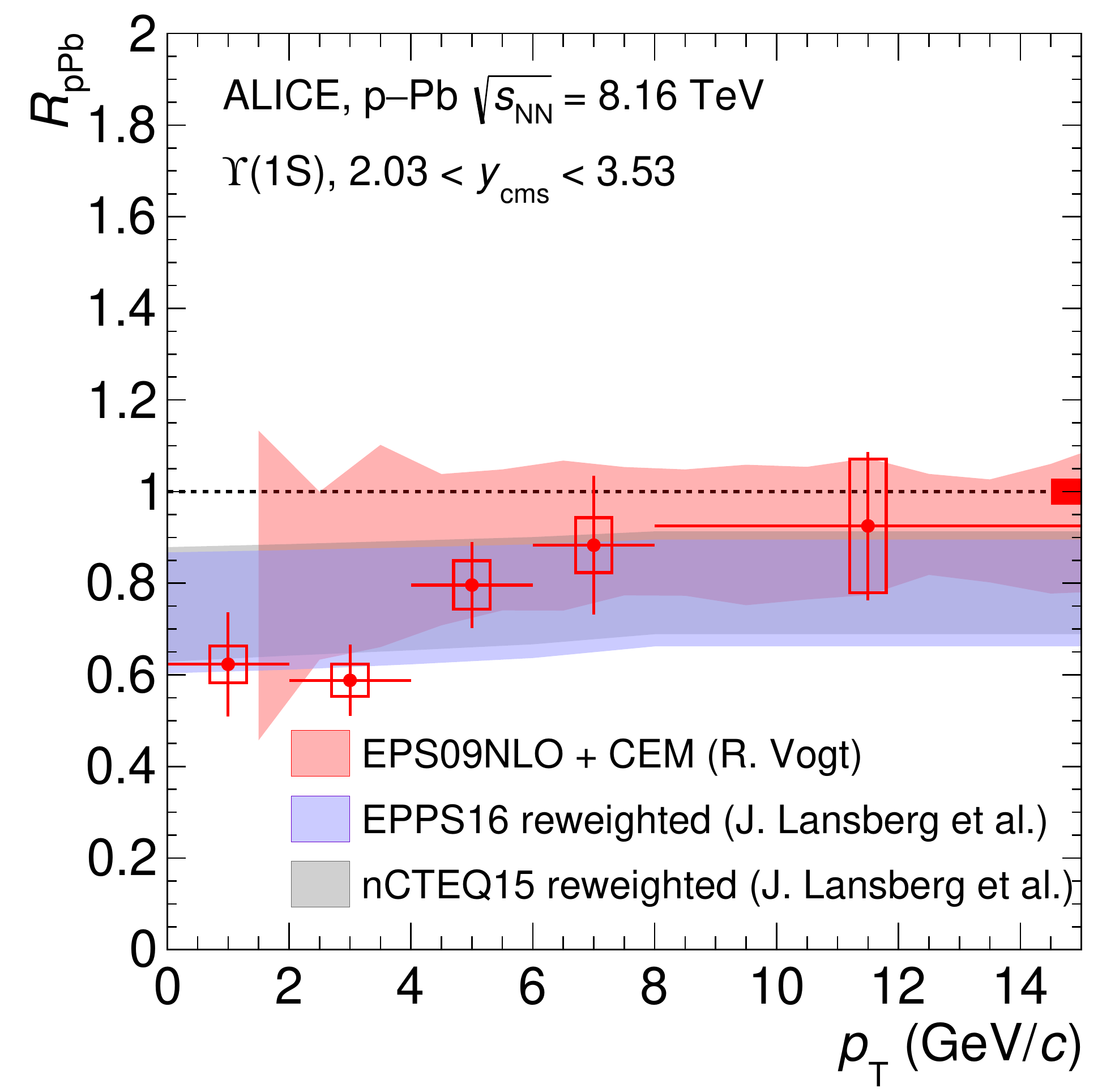}
\caption{\upsone \RpPb as a function of \pt for \Pbp (left panel) and \pPb collisions (right panel). The \RpPb values are compared with  theoretical calculations based on EPS09 NLO~\cite{Vogt:2015uba,Albacete:2017qng}, nCTEQ15 and EPPS16~\cite{Lansberg:2016deg,Kusina:2017gkz,Shao:2015vga,Shao:2012iz,Kovarik:2015cma,Eskola:2016oht} shadowing implementations. 
Details on the theory uncertainty bands are discussed in the text.}
\label{fig:RpA_pt} 
\end{center}
\end{figure}

The \pt dependence of the \upsone \RpPb is shown in Fig.~\ref{fig:RpA_pt}. A slight decrease of the \upsone nuclear modification factor, with decreasing \pt, is observed.  The behaviour is similar both at backward and forward rapidities.

The \ycms and \pt dependence of the \upsone \RpPb are compared, in Fig.~\ref{fig:RpA_y} and Fig.~\ref{fig:RpA_pt}, to several models (referred in the following as nuclear shadowing models), based on EPS09~\cite{Eskola:2009uj}, nCTEQ15~\cite{Kovarik:2015cma} or  EPPS16~\cite{Eskola:2016oht} sets of nuclear parton distribution functions. The EPS09 next-to-leading order (NLO) parametrisation is combined with a  NLO Colour Evaporation Model (CEM)~\cite{Vogt:2015uba}, which describes the \upsi production. The corresponding uncertainty bands, shown in Fig.~\ref{fig:RpA_y} and Fig.~\ref{fig:RpA_pt}, are dominated by the uncertainties of the EPS09 parametrisation. The nCTEQ15 and the EPPS16 NLO nPDFs sets are implemented following the Bayesian reweighting procedure described in ~\cite{Lansberg:2016deg,Kusina:2017gkz,Shao:2015vga,Shao:2012iz}. The uncertainty bands, in this case, represent the convolution of the uncertainties on the nPDFs sets and those on the factorisation scales. 
It can be observed that the shadowing calculations describe fairly well the \pt and \ycms dependence of the \upsone nuclear modification factor in $2.03 < y_{\rm{cms}} < 3.05$, while they overestimate the results obtained in $-4.46 < y_{\rm{cms}} < -2.96$. Furthermore, while the \pt dependence of the ALICE measurements indicate slightly stronger cold nuclear matter effects at low \pt, the shadowing calculations suggest a flatter behaviour. 
Finally, the \ycms dependence of the \RpPb is also compared with a model which includes the effects of parton coherent energy loss with or without the contribution of the EPS09 nuclear shadowing~\cite{Arleo:2012rs,Albacete:2017qng}. The model predicts a mild dependence of the energy loss mechanism on rapidity. When the nuclear shadowing contribution is included, the model describes the forward-rapidity results, while it slightly overestimates the backward-rapidity \RpPb. 
The \upsone \RpPb is also compared with a theoretical model which includes a shadowing contribution, based on the nCTEQ15 set of nPDFs, on top of a suppression of the \upsone production due to interactions with comoving particles~\cite{Ferreiro:2018vmr,Ferreiro:2018wbd}. The uncertainties associated to this theoretical calculation include a small contribution from the uncertainty on the comovers cross section and are dominated by the uncertainties on the shadowing. Also in this case the calculation slightly overestimates the ALICE measurements at  backward \ycms, while at forward \ycms the data agree with the model. It can be noted that the interpretation of the \upsone behaviour in \pPb collisions would also benefit from a precise knowledge, so far still affected by large uncertainties, of the feed-down contribution of the excited states into the \upsone.

\begin{figure}[ht]
\begin{center}
\includegraphics[width=0.49\linewidth,keepaspectratio]{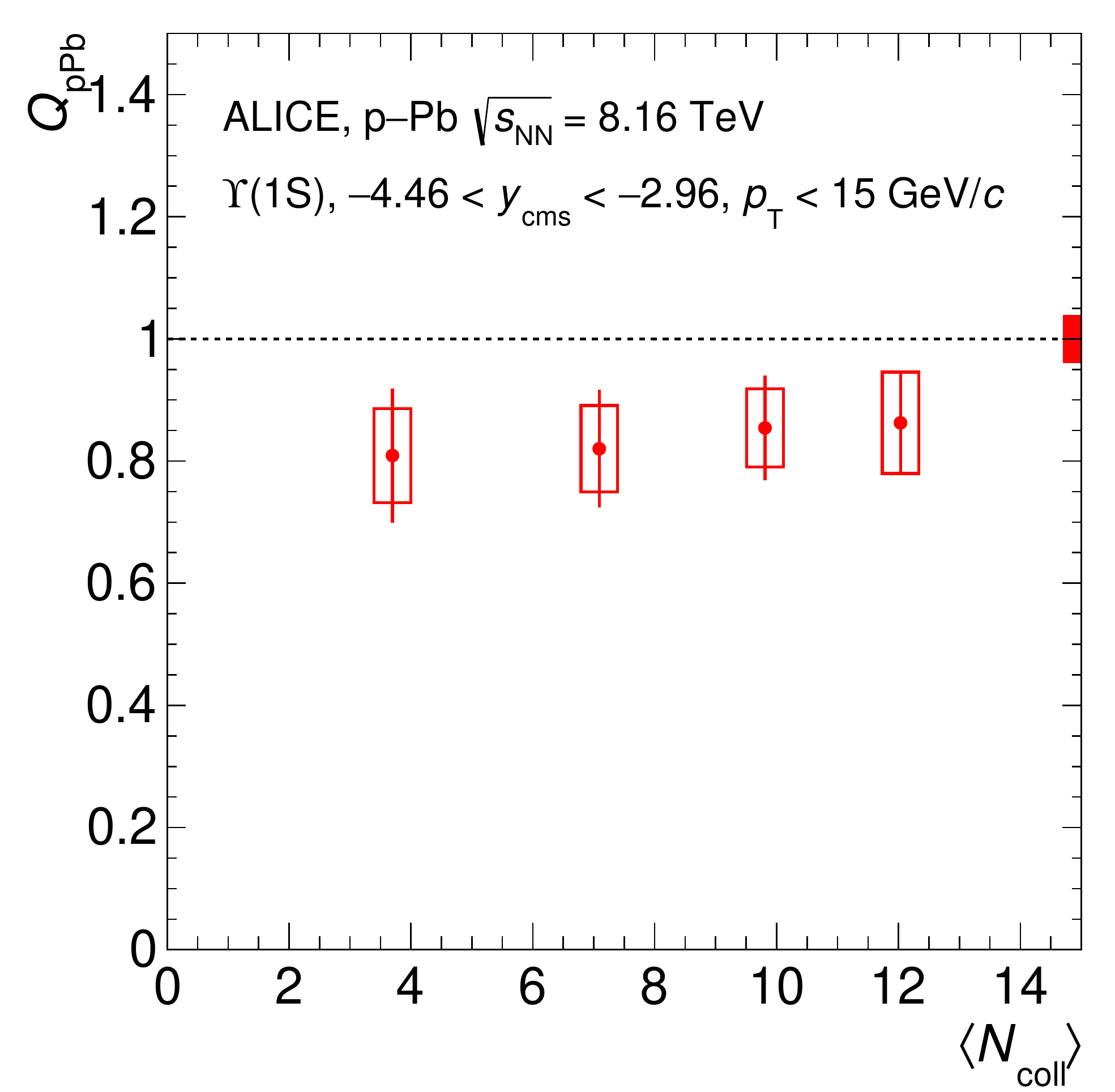}  
\includegraphics[width=0.49\linewidth,keepaspectratio]{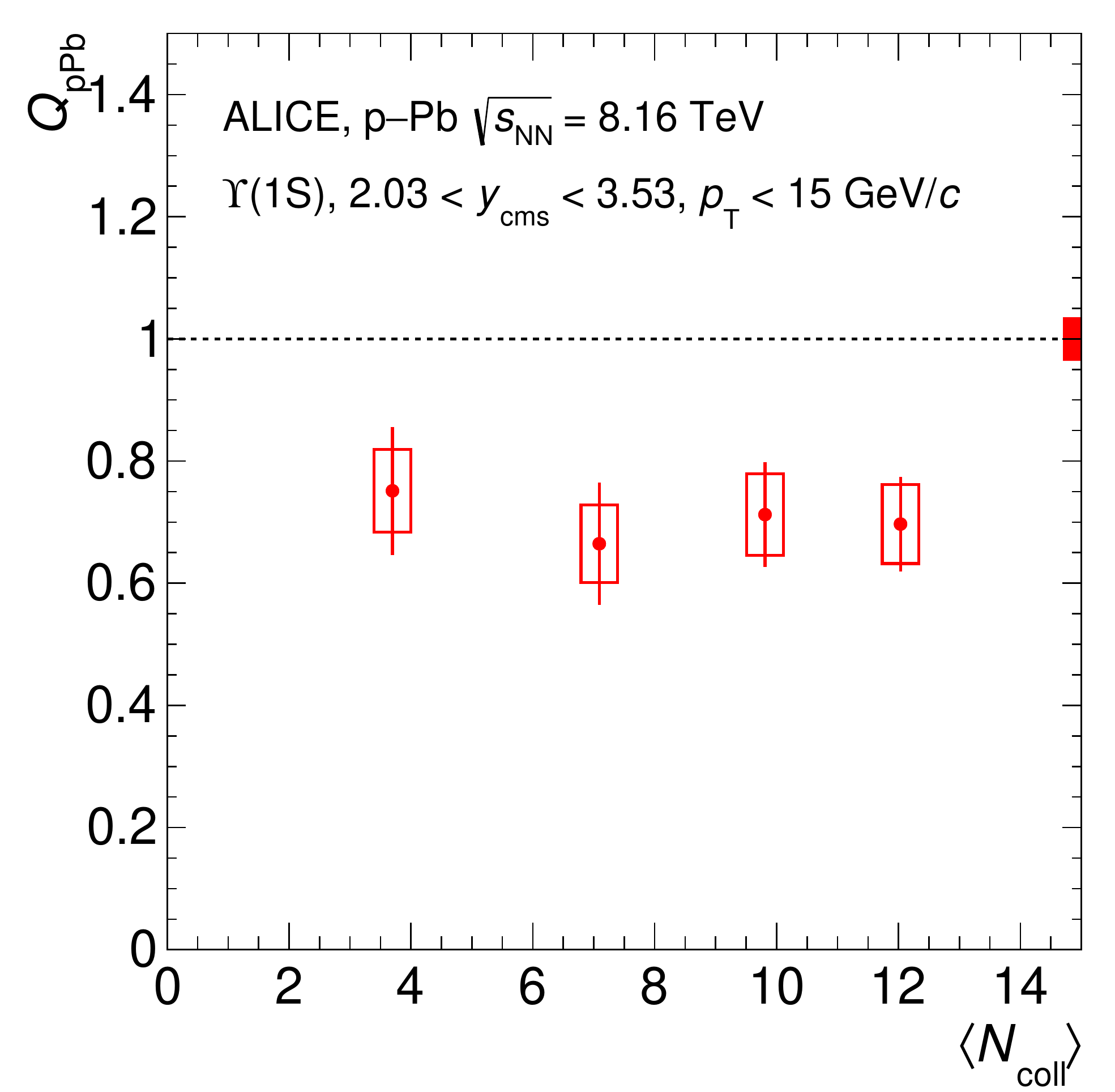}  
\caption{\upsone \QpPb as a function of \avncoll, for \Pbp (left panel) and \pPb collisions (right panel). 
}
\label{fig:QpPb} 
\end{center}
\end{figure}

The \upsone nuclear modification factor is evaluated as a function of the collision centrality. 
The \QpPb results, shown in Fig.~\ref{fig:QpPb}, are presented as a function of the average number of collisions, \avncoll and it can be observed that both at forward and backward rapidity the \upsone centrality dependence is rather flat.

\begin{figure}[ht!]
\begin{center}
\includegraphics[width=0.7\linewidth]{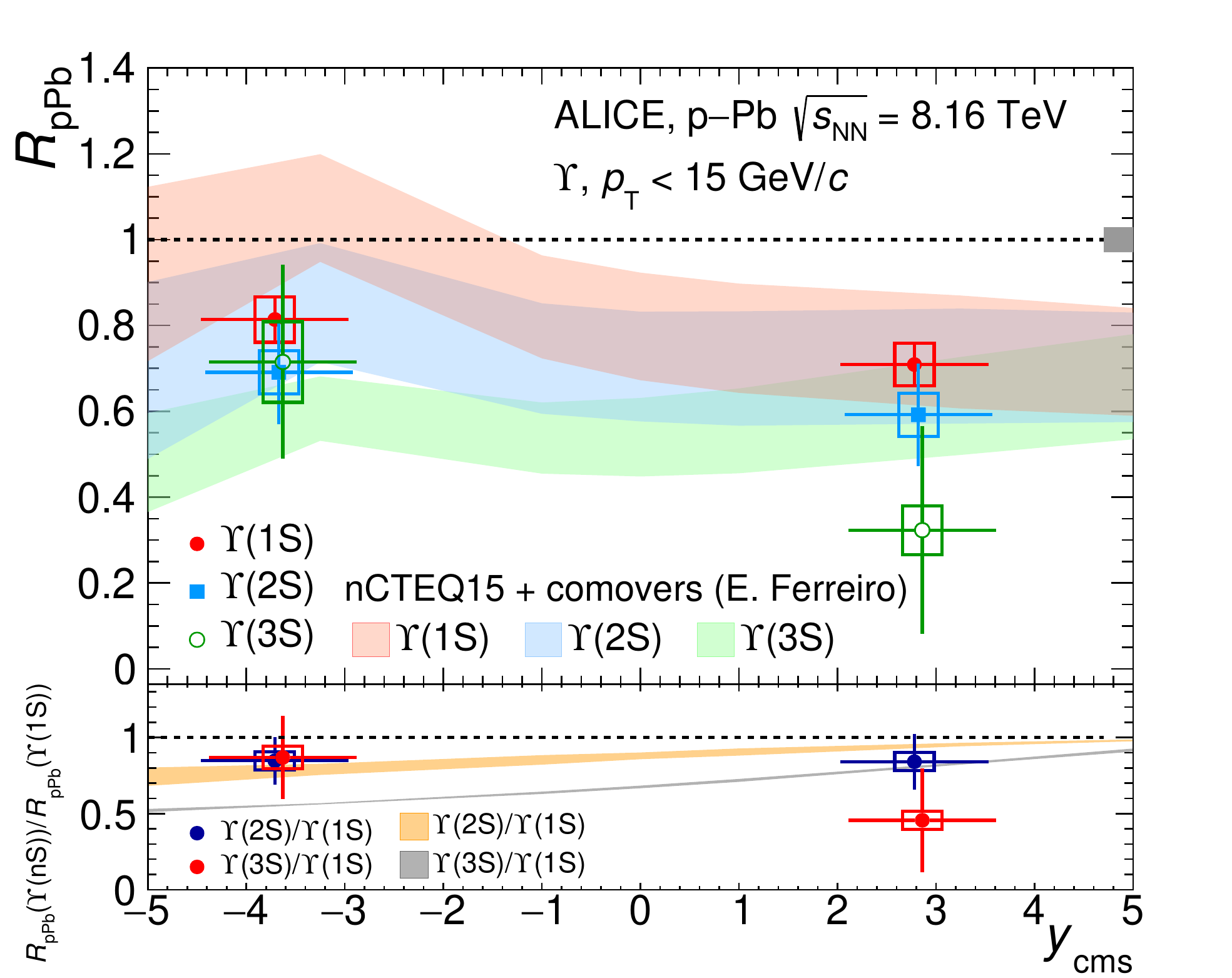}
\caption{\upsone, \upstwo and \upsthree \RpPb at \sqrts as a function of \ycms. 
The \RpPb values of the three resonances are slightly displaced horizontally to improve visibility. Theoretical calculations including nCTEQ15 shadowing contribution and interactions between the \upsi states and comoving particles~\cite{Ferreiro:2018vmr,Ferreiro:2018wbd} are also shown for all the resonances. The grey box around unity represents the global uncertainty common to the three \upsi states. In the lower panel, the ratio of the \upstwo to \upsone and \upsthree to \upsone \RpPb is shown, together with a calculation based on the aforementioned theory model~\cite{Ferreiro:2018vmr,Ferreiro:2018wbd}.}
\label{fig:RpA_2S} 
\end{center}
\end{figure}

Finally, the nuclear modification factor is also evaluated for the \upstwo and \upsthree resonances, in the forward and backward-\ycms intervals, as shown in Fig.~\ref{fig:RpA_2S}. The corresponding \upstwo \RpPb values are:
\begin{align*}
R_{\rm{pPb}}^{\Upsilon\rm{(2S)}} (2.03 < y_{\rm{cms}} < 3.53) = 0.59 \pm 0.12 \;\rm{(stat.)} \pm 0.05 \;\rm{(uncor.\,syst.)} \pm 0.02 \;\rm{(cor.\,syst.)} \\
R_{\rm{pPb}}^{\Upsilon\rm{(2S)}} (-4.46 < y_{\rm{cms}} < -2.96) = 0.69 \pm 0.12 \;\rm{(stat.)} \pm 0.05 \;\rm{(uncor.\,syst.)} \pm 0.02 \;\rm{(cor.\,syst.)} 
\end{align*}
the \upstwo suppression being compatible with unity within 3.1$\sigma$ at forward \ycms and 2.3$\sigma$ at backward \ycms.
The \upsthree \RpPb values are:
\begin{align*}
R_{\rm{pPb}}^{\Upsilon\rm{(3S)}} (2.03 < y_{\rm{cms}} < 3.53) = 0.32 \pm 0.24 \;\rm{(stat.)} \pm 0.06 \;\rm{(uncor.\,syst.)} \pm 0.01 \;\rm{(cor.\,syst.)} \\
R_{\rm{pPb}}^{\Upsilon\rm{(3S)}} (-4.46 < y_{\rm{cms}} < -2.96) = 0.71 \pm 0.23 \;\rm{(stat.)} \pm 0.09 \;\rm{(uncor.\,syst.)} \pm 0.02 \;\rm{(cor.\,syst.)} 
\end{align*}
The \upsthree suppression is compatible with unity within 2.7$\sigma$ at forward \ycms and 1.2$\sigma$ at backward \ycms.
The difference in the \RpPb of the \upstwo and \upsone amounts to 0.5$\sigma$ in both rapidity intervals, suggesting, in \pPb collisions, a similar modification of the production yields of the two \upsi states, with respect to \pp collisions. Unfortunately, the large uncertainties on the \upsthree prevent robust conclusions on the behaviour of the most loosely bound bottomonium state. 
The model which includes both the nuclear shadowing contribution (nCTEQ15) and interactions with comoving particles~\cite{Ferreiro:2018vmr,Ferreiro:2018wbd} suggests a small difference between the nuclear modification factors of the three \upsi states. This difference is slightly more important in the backward-rapidity range, while it becomes negligible at forward \ycms. 
By evaluating the ratio of the $\Upsilon\rm{(nS)}$ to \upsone nuclear modification factors, the shadowing contribution and most of the theory uncertainties, as well as some of the uncertainties on the data, cancel out. The shape of the theoretical calculation is, hence, mainly driven by the interactions with the comoving particles, which affect mostly the excited \upsi states in the backward rapidity region. As shown in the lower panel of Fig.~\ref{fig:RpA_2S}, the ALICE measurements and the model are in fair agreement, even if the uncertainties on the data do not yet allow a firm conclusion on the role of comovers to be drawn.

\section{Conclusions}
The ALICE measurements of the rapidity, transverse momentum and centrality dependence of the inclusive \upsone nuclear modification factor in \pPb collisions at \sqrts have been presented. 
The results show a suppression of the \upsone yields, with respect to the ones measured in \mbox{pp} collisions at the same centre-of-mass energy. The \RpPb values are similar at forward and backward rapidity with a slightly stronger suppression at low \pt, while 
in both rapidity intervals there is no evidence for a centrality dependence of the \upsone \QpPb.
The results obtained at \sqrts are similar within uncertainties to those measured by ALICE in \pPb collisions at the lower energy of \sqrtsfive and show a good agreement with the LHCb measurements at the same centre-of-mass energy. Models based on nuclear shadowing, coherent parton energy loss or interactions with comoving particles fairly describe the data at forward rapidity, while they tend to overestimate the \RpPb at backward \ycms. 
The \upstwo \RpPb has also been measured, showing a strong suppression, similar to the one measured for the \upsone, in the two investigated rapidity intervals. Finally, a first measurement of the \upsthree has also been performed, even if the large uncertainties prevent a detailed comparison of its behaviour in \pPb collisions with respect to the other bottomonium states.
These new bottomonium measurements represent an important baseline for the understanding of the role of CNM effects in \pPb collisions and open up the way for future precision analyses with the upcoming LHC Run 3 and Run 4 data taking periods.

%

\newenvironment{acknowledgement}{\relax}{\relax}
\begin{acknowledgement}
\section*{Acknowledgements}

The ALICE Collaboration would like to thank all its engineers and technicians for their invaluable contributions to the construction of the experiment and the CERN accelerator teams for the outstanding performance of the LHC complex.
The ALICE Collaboration gratefully acknowledges the resources and support provided by all Grid centres and the Worldwide LHC Computing Grid (WLCG) collaboration.
The ALICE Collaboration acknowledges the following funding agencies for their support in building and running the ALICE detector:
A. I. Alikhanyan National Science Laboratory (Yerevan Physics Institute) Foundation (ANSL), State Committee of Science and World Federation of Scientists (WFS), Armenia;
Austrian Academy of Sciences, Austrian Science Fund (FWF): [M 2467-N36] and Nationalstiftung f\"{u}r Forschung, Technologie und Entwicklung, Austria;
Ministry of Communications and High Technologies, National Nuclear Research Center, Azerbaijan;
Conselho Nacional de Desenvolvimento Cient\'{\i}fico e Tecnol\'{o}gico (CNPq), Financiadora de Estudos e Projetos (Finep), Funda\c{c}\~{a}o de Amparo \`{a} Pesquisa do Estado de S\~{a}o Paulo (FAPESP) and Universidade Federal do Rio Grande do Sul (UFRGS), Brazil;
Ministry of Education of China (MOEC) , Ministry of Science \& Technology of China (MSTC) and National Natural Science Foundation of China (NSFC), China;
Ministry of Science and Education and Croatian Science Foundation, Croatia;
Centro de Aplicaciones Tecnol\'{o}gicas y Desarrollo Nuclear (CEADEN), Cubaenerg\'{\i}a, Cuba;
Ministry of Education, Youth and Sports of the Czech Republic, Czech Republic;
The Danish Council for Independent Research | Natural Sciences, the VILLUM FONDEN and Danish National Research Foundation (DNRF), Denmark;
Helsinki Institute of Physics (HIP), Finland;
Commissariat \`{a} l'Energie Atomique (CEA), Institut National de Physique Nucl\'{e}aire et de Physique des Particules (IN2P3) and Centre National de la Recherche Scientifique (CNRS) and R\'{e}gion des  Pays de la Loire, France;
Bundesministerium f\"{u}r Bildung und Forschung (BMBF) and GSI Helmholtzzentrum f\"{u}r Schwerionenforschung GmbH, Germany;
General Secretariat for Research and Technology, Ministry of Education, Research and Religions, Greece;
National Research, Development and Innovation Office, Hungary;
Department of Atomic Energy Government of India (DAE), Department of Science and Technology, Government of India (DST), University Grants Commission, Government of India (UGC) and Council of Scientific and Industrial Research (CSIR), India;
Indonesian Institute of Science, Indonesia;
Centro Fermi - Museo Storico della Fisica e Centro Studi e Ricerche Enrico Fermi and Istituto Nazionale di Fisica Nucleare (INFN), Italy;
Institute for Innovative Science and Technology , Nagasaki Institute of Applied Science (IIST), Japanese Ministry of Education, Culture, Sports, Science and Technology (MEXT) and Japan Society for the Promotion of Science (JSPS) KAKENHI, Japan;
Consejo Nacional de Ciencia (CONACYT) y Tecnolog\'{i}a, through Fondo de Cooperaci\'{o}n Internacional en Ciencia y Tecnolog\'{i}a (FONCICYT) and Direcci\'{o}n General de Asuntos del Personal Academico (DGAPA), Mexico;
Nederlandse Organisatie voor Wetenschappelijk Onderzoek (NWO), Netherlands;
The Research Council of Norway, Norway;
Commission on Science and Technology for Sustainable Development in the South (COMSATS), Pakistan;
Pontificia Universidad Cat\'{o}lica del Per\'{u}, Peru;
Ministry of Science and Higher Education and National Science Centre, Poland;
Korea Institute of Science and Technology Information and National Research Foundation of Korea (NRF), Republic of Korea;
Ministry of Education and Scientific Research, Institute of Atomic Physics and Ministry of Research and Innovation and Institute of Atomic Physics, Romania;
Joint Institute for Nuclear Research (JINR), Ministry of Education and Science of the Russian Federation, National Research Centre Kurchatov Institute, Russian Science Foundation and Russian Foundation for Basic Research, Russia;
Ministry of Education, Science, Research and Sport of the Slovak Republic, Slovakia;
National Research Foundation of South Africa, South Africa;
Swedish Research Council (VR) and Knut \& Alice Wallenberg Foundation (KAW), Sweden;
European Organization for Nuclear Research, Switzerland;
Suranaree University of Technology (SUT), National Science and Technology Development Agency (NSDTA) and Office of the Higher Education Commission under NRU project of Thailand, Thailand;
Turkish Atomic Energy Agency (TAEK), Turkey;
National Academy of  Sciences of Ukraine, Ukraine;
Science and Technology Facilities Council (STFC), United Kingdom;
National Science Foundation of the United States of America (NSF) and United States Department of Energy, Office of Nuclear Physics (DOE NP), United States of America.    
\end{acknowledgement}
%
\bibliographystyle{utphys}
\bibliography{Paper_Upsilon_v8}{}
\newpage
\appendix
\section{The ALICE Collaboration}
\label{app:collab}

\begingroup
\small
\begin{flushleft}
S.~Acharya\Irefn{org141}\And 
D.~Adamov\'{a}\Irefn{org94}\And 
A.~Adler\Irefn{org74}\And 
J.~Adolfsson\Irefn{org80}\And 
M.M.~Aggarwal\Irefn{org99}\And 
G.~Aglieri Rinella\Irefn{org33}\And 
M.~Agnello\Irefn{org30}\And 
N.~Agrawal\Irefn{org10}\textsuperscript{,}\Irefn{org53}\And 
Z.~Ahammed\Irefn{org141}\And 
S.~Ahmad\Irefn{org16}
% ...
\And
\renewcommand\labelenumi{\textsuperscript{\theenumi}~}

\section*{Affiliation notes}
\renewcommand\theenumi{\roman{enumi}}
\begin{Authlist}
\item \Adef{org*}Deceased
\item \Adef{orgI}Dipartimento DET del Politecnico di Torino, Turin, Italy
\item \Adef{orgII}M.V. Lomonosov Moscow State University, D.V. Skobeltsyn Institute of Nuclear, Physics, Moscow, Russia
\item \Adef{orgIII}Department of Applied Physics, Aligarh Muslim University, Aligarh, India
\item \Adef{orgIV}Institute of Theoretical Physics, University of Wroclaw, Poland
\end{Authlist}

\section*{Collaboration Institutes}
\renewcommand\theenumi{\arabic{enumi}~}
\begin{Authlist}
\item \Idef{org1}A.I. Alikhanyan National Science Laboratory (Yerevan Physics Institute) Foundation, Yerevan, Armenia
\item \Idef{org2}Bogolyubov Institute for Theoretical Physics, National Academy of Sciences of Ukraine, Kiev, Ukraine
\item \Idef{org3}Bose Institute, Department of Physics  and Centre for Astroparticle Physics and Space Science (CAPSS), Kolkata, India
\item \Idef{org4}Budker Institute for Nuclear Physics, Novosibirsk, Russia
\item \Idef{org5}California Polytechnic State University, San Luis Obispo, California, United States
\item \Idef{org6}Central China Normal University, Wuhan, China
\item \Idef{org7}Centre de Calcul de l'IN2P3, Villeurbanne, Lyon, France
\item \Idef{org8}Centro de Aplicaciones Tecnol\'{o}gicas y Desarrollo Nuclear (CEADEN), Havana, Cuba
\item \Idef{org9}Centro de Investigaci\'{o}n y de Estudios Avanzados (CINVESTAV), Mexico City and M\'{e}rida, Mexico
\item \Idef{org10}Centro Fermi - Museo Storico della Fisica e Centro Studi e Ricerche ``Enrico Fermi', Rome, Italy
\item \Idef{org11}Chicago State University, Chicago, Illinois, United States
\item \Idef{org12}China Institute of Atomic Energy, Beijing, China
\item \Idef{org13}Comenius University Bratislava, Faculty of Mathematics, Physics and Informatics, Bratislava, Slovakia
\item \Idef{org14}COMSATS University Islamabad, Islamabad, Pakistan
\item \Idef{org15}Creighton University, Omaha, Nebraska, United States
\item \Idef{org16}Department of Physics, Aligarh Muslim University, Aligarh, India
\item \Idef{org17}Department of Physics, Pusan National University, Pusan, Republic of Korea
\item \Idef{org18}Department of Physics, Sejong University, Seoul, Republic of Korea
\item \Idef{org19}Department of Physics, University of California, Berkeley, California, United States
\item \Idef{org20}Department of Physics, University of Oslo, Oslo, Norway
\item \Idef{org21}Department of Physics and Technology, University of Bergen, Bergen, Norway
\item \Idef{org22}Dipartimento di Fisica dell'Universit\`{a} 'La Sapienza' and Sezione INFN, Rome, Italy
\item \Idef{org23}Dipartimento di Fisica dell'Universit\`{a} and Sezione INFN, Cagliari, Italy
\item \Idef{org24}Dipartimento di Fisica dell'Universit\`{a} and Sezione INFN, Trieste, Italy
\item \Idef{org25}Dipartimento di Fisica dell'Universit\`{a} and Sezione INFN, Turin, Italy
\item \Idef{org26}Dipartimento di Fisica e Astronomia dell'Universit\`{a} and Sezione INFN, Bologna, Italy
\item \Idef{org27}Dipartimento di Fisica e Astronomia dell'Universit\`{a} and Sezione INFN, Catania, Italy
\item \Idef{org28}Dipartimento di Fisica e Astronomia dell'Universit\`{a} and Sezione INFN, Padova, Italy
\item \Idef{org29}Dipartimento di Fisica `E.R.~Caianiello' dell'Universit\`{a} and Gruppo Collegato INFN, Salerno, Italy
\item \Idef{org30}Dipartimento DISAT del Politecnico and Sezione INFN, Turin, Italy
\item \Idef{org31}Dipartimento di Scienze e Innovazione Tecnologica dell'Universit\`{a} del Piemonte Orientale and INFN Sezione di Torino, Alessandria, Italy
\item \Idef{org32}Dipartimento Interateneo di Fisica `M.~Merlin' and Sezione INFN, Bari, Italy
\item \Idef{org33}European Organization for Nuclear Research (CERN), Geneva, Switzerland
\item \Idef{org34}Faculty of Electrical Engineering, Mechanical Engineering and Naval Architecture, University of Split, Split, Croatia
\item \Idef{org35}Faculty of Engineering and Science, Western Norway University of Applied Sciences, Bergen, Norway
\item \Idef{org36}Faculty of Nuclear Sciences and Physical Engineering, Czech Technical University in Prague, Prague, Czech Republic
\item \Idef{org37}Faculty of Science, P.J.~\v{S}af\'{a}rik University, Ko\v{s}ice, Slovakia
\item \Idef{org38}Frankfurt Institute for Advanced Studies, Johann Wolfgang Goethe-Universit\"{a}t Frankfurt, Frankfurt, Germany
\item \Idef{org39}Fudan University, Shanghai, China
\item \Idef{org40}Gangneung-Wonju National University, Gangneung, Republic of Korea
\item \Idef{org41}Gauhati University, Department of Physics, Guwahati, India
\item \Idef{org42}Helmholtz-Institut f\"{u}r Strahlen- und Kernphysik, Rheinische Friedrich-Wilhelms-Universit\"{a}t Bonn, Bonn, Germany
\item \Idef{org43}Helsinki Institute of Physics (HIP), Helsinki, Finland
\item \Idef{org44}High Energy Physics Group,  Universidad Aut\'{o}noma de Puebla, Puebla, Mexico
\item \Idef{org45}Hiroshima University, Hiroshima, Japan
\item \Idef{org46}Hochschule Worms, Zentrum  f\"{u}r Technologietransfer und Telekommunikation (ZTT), Worms, Germany
\item \Idef{org47}Horia Hulubei National Institute of Physics and Nuclear Engineering, Bucharest, Romania
\item \Idef{org48}Indian Institute of Technology Bombay (IIT), Mumbai, India
\item \Idef{org49}Indian Institute of Technology Indore, Indore, India
\item \Idef{org50}Indonesian Institute of Sciences, Jakarta, Indonesia
\item \Idef{org51}INFN, Laboratori Nazionali di Frascati, Frascati, Italy
\item \Idef{org52}INFN, Sezione di Bari, Bari, Italy
\item \Idef{org53}INFN, Sezione di Bologna, Bologna, Italy
\item \Idef{org54}INFN, Sezione di Cagliari, Cagliari, Italy
\item \Idef{org55}INFN, Sezione di Catania, Catania, Italy
\item \Idef{org56}INFN, Sezione di Padova, Padova, Italy
\item \Idef{org57}INFN, Sezione di Roma, Rome, Italy
\item \Idef{org58}INFN, Sezione di Torino, Turin, Italy
\item \Idef{org59}INFN, Sezione di Trieste, Trieste, Italy
\item \Idef{org60}Inha University, Incheon, Republic of Korea
\item \Idef{org61}Institut de Physique Nucl\'{e}aire d'Orsay (IPNO), Institut National de Physique Nucl\'{e}aire et de Physique des Particules (IN2P3/CNRS), Universit\'{e} de Paris-Sud, Universit\'{e} Paris-Saclay, Orsay, France
\item \Idef{org62}Institute for Nuclear Research, Academy of Sciences, Moscow, Russia
\item \Idef{org63}Institute for Subatomic Physics, Utrecht University/Nikhef, Utrecht, Netherlands
\item \Idef{org64}Institute of Experimental Physics, Slovak Academy of Sciences, Ko\v{s}ice, Slovakia
\item \Idef{org65}Institute of Physics, Homi Bhabha National Institute, Bhubaneswar, India
\item \Idef{org66}Institute of Physics of the Czech Academy of Sciences, Prague, Czech Republic
\item \Idef{org67}Institute of Space Science (ISS), Bucharest, Romania
\item \Idef{org68}Institut f\"{u}r Kernphysik, Johann Wolfgang Goethe-Universit\"{a}t Frankfurt, Frankfurt, Germany
\item \Idef{org69}Instituto de Ciencias Nucleares, Universidad Nacional Aut\'{o}noma de M\'{e}xico, Mexico City, Mexico
\item \Idef{org70}Instituto de F\'{i}sica, Universidade Federal do Rio Grande do Sul (UFRGS), Porto Alegre, Brazil
\item \Idef{org71}Instituto de F\'{\i}sica, Universidad Nacional Aut\'{o}noma de M\'{e}xico, Mexico City, Mexico
\item \Idef{org72}iThemba LABS, National Research Foundation, Somerset West, South Africa
\item \Idef{org73}Jeonbuk National University, Jeonju, Republic of Korea
\item \Idef{org74}Johann-Wolfgang-Goethe Universit\"{a}t Frankfurt Institut f\"{u}r Informatik, Fachbereich Informatik und Mathematik, Frankfurt, Germany
\item \Idef{org75}Joint Institute for Nuclear Research (JINR), Dubna, Russia
\item \Idef{org76}Korea Institute of Science and Technology Information, Daejeon, Republic of Korea
\item \Idef{org77}KTO Karatay University, Konya, Turkey
\item \Idef{org78}Laboratoire de Physique Subatomique et de Cosmologie, Universit\'{e} Grenoble-Alpes, CNRS-IN2P3, Grenoble, France
\item \Idef{org79}Lawrence Berkeley National Laboratory, Berkeley, California, United States
\item \Idef{org80}Lund University Department of Physics, Division of Particle Physics, Lund, Sweden
\item \Idef{org81}Nagasaki Institute of Applied Science, Nagasaki, Japan
\item \Idef{org82}Nara Women{'}s University (NWU), Nara, Japan
\item \Idef{org83}National and Kapodistrian University of Athens, School of Science, Department of Physics , Athens, Greece
\item \Idef{org84}National Centre for Nuclear Research, Warsaw, Poland
\item \Idef{org85}National Institute of Science Education and Research, Homi Bhabha National Institute, Jatni, India
\item \Idef{org86}National Nuclear Research Center, Baku, Azerbaijan
\item \Idef{org87}National Research Centre Kurchatov Institute, Moscow, Russia
\item \Idef{org88}Niels Bohr Institute, University of Copenhagen, Copenhagen, Denmark
\item \Idef{org89}Nikhef, National institute for subatomic physics, Amsterdam, Netherlands
\item \Idef{org90}NRC Kurchatov Institute IHEP, Protvino, Russia
\item \Idef{org91}NRC «Kurchatov Institute»  - ITEP, Moscow, Russia
\item \Idef{org92}NRNU Moscow Engineering Physics Institute, Moscow, Russia
\item \Idef{org93}Nuclear Physics Group, STFC Daresbury Laboratory, Daresbury, United Kingdom
\item \Idef{org94}Nuclear Physics Institute of the Czech Academy of Sciences, \v{R}e\v{z} u Prahy, Czech Republic
\item \Idef{org95}Oak Ridge National Laboratory, Oak Ridge, Tennessee, United States
\item \Idef{org96}Ohio State University, Columbus, Ohio, United States
\item \Idef{org97}Petersburg Nuclear Physics Institute, Gatchina, Russia
\item \Idef{org98}Physics department, Faculty of science, University of Zagreb, Zagreb, Croatia
\item \Idef{org99}Physics Department, Panjab University, Chandigarh, India
\item \Idef{org100}Physics Department, University of Jammu, Jammu, India
\item \Idef{org101}Physics Department, University of Rajasthan, Jaipur, India
\item \Idef{org102}Physikalisches Institut, Eberhard-Karls-Universit\"{a}t T\"{u}bingen, T\"{u}bingen, Germany
\item \Idef{org103}Physikalisches Institut, Ruprecht-Karls-Universit\"{a}t Heidelberg, Heidelberg, Germany
\item \Idef{org104}Physik Department, Technische Universit\"{a}t M\"{u}nchen, Munich, Germany
\item \Idef{org105}Politecnico di Bari, Bari, Italy
\item \Idef{org106}Research Division and ExtreMe Matter Institute EMMI, GSI Helmholtzzentrum f\"ur Schwerionenforschung GmbH, Darmstadt, Germany
\item \Idef{org107}Rudjer Bo\v{s}kovi\'{c} Institute, Zagreb, Croatia
\item \Idef{org108}Russian Federal Nuclear Center (VNIIEF), Sarov, Russia
\item \Idef{org109}Saha Institute of Nuclear Physics, Homi Bhabha National Institute, Kolkata, India
\item \Idef{org110}School of Physics and Astronomy, University of Birmingham, Birmingham, United Kingdom
\item \Idef{org111}Secci\'{o}n F\'{\i}sica, Departamento de Ciencias, Pontificia Universidad Cat\'{o}lica del Per\'{u}, Lima, Peru
\item \Idef{org112}St. Petersburg State University, St. Petersburg, Russia
\item \Idef{org113}Stefan Meyer Institut f\"{u}r Subatomare Physik (SMI), Vienna, Austria
\item \Idef{org114}SUBATECH, IMT Atlantique, Universit\'{e} de Nantes, CNRS-IN2P3, Nantes, France
\item \Idef{org115}Suranaree University of Technology, Nakhon Ratchasima, Thailand
\item \Idef{org116}Technical University of Ko\v{s}ice, Ko\v{s}ice, Slovakia
\item \Idef{org117}Technische Universit\"{a}t M\"{u}nchen, Excellence Cluster 'Universe', Munich, Germany
\item \Idef{org118}The Henryk Niewodniczanski Institute of Nuclear Physics, Polish Academy of Sciences, Cracow, Poland
\item \Idef{org119}The University of Texas at Austin, Austin, Texas, United States
\item \Idef{org120}Universidad Aut\'{o}noma de Sinaloa, Culiac\'{a}n, Mexico
\item \Idef{org121}Universidade de S\~{a}o Paulo (USP), S\~{a}o Paulo, Brazil
\item \Idef{org122}Universidade Estadual de Campinas (UNICAMP), Campinas, Brazil
\item \Idef{org123}Universidade Federal do ABC, Santo Andre, Brazil
\item \Idef{org124}University of Cape Town, Cape Town, South Africa
\item \Idef{org125}University of Houston, Houston, Texas, United States
\item \Idef{org126}University of Jyv\"{a}skyl\"{a}, Jyv\"{a}skyl\"{a}, Finland
\item \Idef{org127}University of Liverpool, Liverpool, United Kingdom
\item \Idef{org128}University of Science and Technology of China, Hefei, China
\item \Idef{org129}University of South-Eastern Norway, Tonsberg, Norway
\item \Idef{org130}University of Tennessee, Knoxville, Tennessee, United States
\item \Idef{org131}University of the Witwatersrand, Johannesburg, South Africa
\item \Idef{org132}University of Tokyo, Tokyo, Japan
\item \Idef{org133}University of Tsukuba, Tsukuba, Japan
\item \Idef{org134}Universit\'{e} Clermont Auvergne, CNRS/IN2P3, LPC, Clermont-Ferrand, France
\item \Idef{org135}Universit\'{e} de Lyon, Universit\'{e} Lyon 1, CNRS/IN2P3, IPN-Lyon, Villeurbanne, Lyon, France
\item \Idef{org136}Universit\'{e} de Strasbourg, CNRS, IPHC UMR 7178, F-67000 Strasbourg, France, Strasbourg, France
\item \Idef{org137}Universit\'{e} Paris-Saclay Centre d'Etudes de Saclay (CEA), IRFU, D\'{e}partment de Physique Nucl\'{e}aire (DPhN), Saclay, France
\item \Idef{org138}Universit\`{a} degli Studi di Foggia, Foggia, Italy
\item \Idef{org139}Universit\`{a} degli Studi di Pavia, Pavia, Italy
\item \Idef{org140}Universit\`{a} di Brescia, Brescia, Italy
\item \Idef{org141}Variable Energy Cyclotron Centre, Homi Bhabha National Institute, Kolkata, India
\item \Idef{org142}Warsaw University of Technology, Warsaw, Poland
\item \Idef{org143}Wayne State University, Detroit, Michigan, United States
\item \Idef{org144}Westf\"{a}lische Wilhelms-Universit\"{a}t M\"{u}nster, Institut f\"{u}r Kernphysik, M\"{u}nster, Germany
\item \Idef{org145}Wigner Research Centre for Physics, Budapest, Hungary
\item \Idef{org146}Yale University, New Haven, Connecticut, United States
\item \Idef{org147}Yonsei University, Seoul, Republic of Korea
\end{Authlist}
\endgroup
%
%
\end{document}